\documentstyle [12pt] {article}

\newcommand{\gtwid}{\mathrel{\raise.3ex\hbox{$>$\kern-.75em\lower1ex
\hbox{$\sim$}}}}
\newcommand{\ltwid}{\mathrel{\raise.3ex\hbox{$<$\kern-.75em\lower1ex
\hbox{$\sim$}}}}
\newcommand{\beq}{\begin{equation}}
\newcommand{\eeq}{\end{equation}}
\newcommand{\beqs}{\begin{eqnarray}}
\newcommand{\eeqs}{\end{eqnarray}}

\newcommand{\prd}{Phys. Rev. D}
\newcommand{\npb}{Nucl. Phys. B}
\newcommand{\plb}{Phys. Lett. B}

\catcode`@=11
\@addtoreset{equation}{section}
\@addtoreset{equation}{subsection}
\def\theequation{\ifnum\value{section}=0 \arabic{equation}\ignorespaces
\else \ifnum\value{section}=-1 A.\arabic{equation}\ignorespaces 
\else \ifnum\value{subsection}=0 \thesection.\arabic{equation}\ignorespaces
\else \thesection.\arabic{subsection}.\arabic{equation}\ignorespaces
                           \fi
                      \fi
                 \fi}
\catcode`@=12

\begin{document}

\def\thefootnote{\fnsymbol{footnote}}
\baselineskip 7.5mm

\begin{flushright}
\begin{tabular}{l}
ITP-SB-93-62    \\
hep-ph/9401274  \\
January, 1994  
\end{tabular}
\end{flushright}

\vspace{8mm}
\begin{center}
{\Large \bf General Analysis of Phases in Quark Mass Matrices  }

\vspace{4mm}
\vspace{16mm}

\setcounter{footnote}{0}
Alexander Kusenko\footnote{email: sasha@max.physics.sunysb.edu}
\setcounter{footnote}{6}
and Robert Shrock\footnote{email: shrock@max.physics.sunysb.edu}

\vspace{6mm}
Institute for Theoretical Physics  \\
State University of New York       \\
Stony Brook, N. Y. 11794-3840  \\

\vspace{20mm}

{\bf Abstract}
\end{center}

We give a detailed discussion of our general determination of (i) the number 
of unremovable, physically meaningful phases in quark mass matrices and (ii)
which elements of these matrices can be rendered real by rephasings of fermion
fields.  The results are applied to several currently viable models.  New
results are presented for an arbitrary number of fermion generations; these
provide further insight into the three-generation case of physical interest. 

\vspace{35mm}

\pagestyle{empty}
\newpage

\pagestyle{plain}
\pagenumbering{arabic}
\renewcommand{\thefootnote}{\arabic{footnote}}
\setcounter{footnote}{0}

\section{ Introduction}
\label{intro}

   Understanding fermion masses and quark mixing remains one of
the most important outstanding problems in particle physics.  In an effort to
gain insight into this problem, many studies of simple models of quark mass
matrices have been carried out over the years.  The phases in these mass 
matrices play an essential role in the Kobayashi-Maskawa (KM) 
mechanism~\cite{km} for CP violation\footnote{As will be discussed further 
below, there are, in general, more
unremovable phases in the quark mass matrices than in the quark mixing matrix,
so that some of the former phases contribute to CP-conserving quantities.
We also note that our analysis does not assume that the KM mechanism is the 
only source of CP violation.}. A given model is characterized by
the number of parameters (amplitudes and phases) which specify the quark mass
matrices. Thus, a very important problem is to determine, for any model, 
how many unremovable, and hence physically meaningful, phases occur in the 
quark mass matrices and which elements of these matrices 
can be made real by rephasings of quark fields.  We recently presented a 
general solution to this problem~\cite{ks}.  Here we give a detailed discussion
of our results and proofs, and applications to currently viable 
models.  To make the paper self-contained, we will review  
the results in \cite{ks}.  In passing, it should be mentioned that we have 
also given a general solution to the analogous problem for lepton mass
matrices\cite{ksl}%
\footnote{The situation in the leptonic sector is qualitatively
more complicated than that in the quark sector because of the general 
presence of three types of fermion bilinears, Dirac, left-handed Majorana, 
and right-handed Majorana, each with its own gauge and rephasing properties. 
It is also different because one does not know the full set of fields which 
might be present: there might or might not exist electroweak singlet 
neutrinos. The most general case was treated in Ref. \cite{ksl}.  The
uncertainty in the field content of the neutrino sector produces the same
uncertainty in a general analysis of phases for both the quark and lepton 
sectors considered together.  An analysis of invariants and their 
constraints in a particular unified model was also given in Ref. 
\cite{ksl}.}.   The organization of this paper is as follows.
In section 2 we discuss our theorem on the number of unremovable phases.
In section 3 we give the details of our complex rephasing invariants and 
results on which elements of the mass matrices can be rendered real.  This 
section contains the proofs of certain theorems presented in Ref. \cite{ks}. 
Section 4 contains some new results on complex invariants 
and unremovable phases for arbitrary $N_G$.  In section 5 we apply our general 
results to models.

\section{ Theorem on the Number of Physical Phases}
\label{number}

     The quark mass terms are taken to arise from interactions which are 
invariant under the standard model gauge group $G_{SM}= SU(3) \times SU(2) 
\times U(1)$, via the spontaneous symmetry breaking of $G_{SM}$.  In the
standard model and its supersymmetric extensions, the resultant mass terms 
appear at the electroweak level via (renormalizable, dimension-4) Yukawa 
couplings$^{3,4}$\addtocounter{footnote}{1}\footnotetext{We do not consider 
models in which fermion masses 
arise via multifermion operators, which are not perturbatively 
renormalizable.  Note that at a nonperturbative level, lattice studies 
\cite{yr} show that a lattice theory with a specific multifermion action 
and no scalar fields may yield the same continuum limit as a theory with 
a Yukawa interaction.}\addtocounter{footnote}{1}%
\footnotetext{In a number of interesting models, 
these Yukawa couplings are viewed as originating at a higher mass scale, 
such as that of a hypothetical supersymmetric grand unified theory (SUSY 
GUT) or some other (supersymmetric) theory resulting from the 
$E << (\alpha')^{-1/2}$ limit of a string
theory. It is possible that interactions which appear to 
be Yukawa couplings at a given mass scale, could be effective, in the sense 
that some of the elements of the associated Yukawa matrices could actually 
arise from higher-dimension operators at a yet higher mass scale.  
An example of a higher-dimension operator which might plausibly occur near 
the scale of quantum gravity is a dimension-5 
operator of the generic form $(c/{\bar M_P})\phi_1 \phi_2 \bar Q_{_L} q_{_R}$,
where $\phi_1$ and $\phi_2$ are appropriate Higgs fields, $c$ denotes a 
dimensionless coefficient, and $\bar M_{P}$ is the (reduced) Planck mass, 
$\bar M_P = \sqrt{\hbar c/(8 \pi G_N)}$.  The plausibility of such
higher-dimension operators can be inferred either from the nonrenormalizability
of supergravity or as a consequence of the $E << (\alpha')^{-1/2}$ 
limit of a string theory.  At lower mass scales this operator could 
produce effective Yukawa terms contributing to (\ref{massterm}), where,  
say, $\phi_1$ contains $H_i$, $i=1$ or 2 as defined in (\ref{h1vev}), 
(\ref{h2vev}), and the factor $c <\phi_2>/{\bar M_P}$ enters into the 
effective Yukawa or mass matrix (where $<\phi_2>$ might have a 
value somewhat smaller than $\bar M_P$ but much larger than the electroweak 
scale).}. It follows that 
these mass terms can be written in terms of the $G_{SM}$ quark fields as 
\beq
-{\cal L}_m = \sum_{j,k=1}^{N_G}
\Bigl [(\bar Q_{j L})_1 M_{jk}^{(u)} u_{k R} 
+ (\bar Q_{j L})_2 M_{jk}^{(d)} d_{k R} \Bigr ] + h.c. 
\label{massterm}
\eeq
where $j$ and $k$ are generation labels; $N_G=3$ is the number of generations 
of standard-model fermions; $Q_{jL}$ is an SU(2) doublet, with 
$Q_{1 L} = (^u_d)_L$, $Q_{2 L} = (^c_s)_L$, $Q_{3 L} = (^t_b)_L$; the 
subscript $a$ on $(Q_{j L})_a$ is the SU(2) index; and the SU(2)-singlet
right-handed quark fields are denoted as $u_{kR}$ with $u_{1 R} = u_R$, 
$u_{2 R} = c_R$, $u_{3 R}=t_R$, $d_{1 R}=d_R$, $d_{2 R}=s_R$, $d_{3 R}=b_R$.
As indicated, we will concentrate on the physical case of $N_G=3$ 
generations of standard-model fermions (with associated light neutrinos).
However, for the sake of generality, and because it provides further insight
into the physical $N_G=3$ case, we will give a number of results for arbitrary 
$N_G$.  $M^{(u)}$ and $M^{(d)}$ are the mass matrices in the up and 
down sectors, whose diagonalization yields the mass eigenstates $u_{jm}$ and 
$d_{jm}$.  Because of the assumed origin of the quark mass matrices from
Yukawa interactions (i.e. interactions which appear as dimension-4 at a given
mass scale, but which could include the contributions of higher-dimension
operators at higher mass scales), it is convenient to write these mass 
matrices in terms of dimensionless Yukawa matrices according to 
\beq
Y^{(f)}=2^{1/2}v_f^{-1}M^{(f)} \ , \quad f=u,d
\label{ym}
\eeq
where $v_u$ and $v_d$ are real quantities with the dimensions of mass, 
and represent Higgs vacuum expectation values in the standard model 
and its supersymmetic extensions.  
For example, in the minimal supersymmetric standard model (MSSM), 
\beq
<H_1> = \left (\begin{array}{c}
                  0 \\
                  2^{-1/2}v_d \end{array}   \right  )
\label{h1vev}
\eeq
and 
\beq
<H_2> = \left (\begin{array}{c}
                  2^{-1/2}v_u \\
                  0  \end{array}   \right  )
\label{h2vev}
\eeq
where $H_1$ and $H_2$ are the $Y=1$ and $Y=-1$ Higgs fields.  With one
definition for the angle $\beta$ in the MSSM, 
$v_u=v \sin \beta$, $v_d=v \cos \beta$, 
where $v \equiv 2^{-1/4}G_F^{-1/2}=246$ GeV (many authors use a different
convention according to which $v_u=v \cos \beta$, $v_d=v \sin \beta$). 
Because $v_u$ and $v_d$ are real, the phase properties of the mass 
matrices $M^{(f)}$ are the same as those of the Yukawa matrices $Y^{(f)}$ 
and we shall deal with them interchangeably.$^{5}$\addtocounter{footnote}{1}%
\footnotetext{In specific 
models in which the Yukawa couplings arise at a high mass scale, e.g. in 
a hypothetical grand unified theory or an effective pointlike field theory 
arising from the $E << (\alpha')^{-1/2}$ limit of a string theory, and in which
below this scale the gauge group reduces to $G_{SM}$ without further 
intermediate structure, it is the $Y^{(f)}$ which one evolves down to the
electroweak scale using the appropriate renormalization group equations.  
Note that $M^{(f)}$ and $Y^{(f)}$ as defined above may subsume the effects 
of several different types of Yukawa couplings of fermion and Higgs 
representations in the theory at the high mass scale, e.g. a GUT.}

    To count the number of unremovable, and hence physically meaningful, 
phases in the quark mass matrices, we rephase the fermion fields in 
(\ref{massterm}) so as to remove all possible phases in these matrices. 

A general theorem on this counting problem was given
before~\cite{ks}; we review it here.  As was evident in Ref. \cite{ks}, this
theorem applies for arbitrary $N_G$; although the physically interesting case
is $N_G=3$, we indicate the general $N_G$-dependence here. 
One can perform the rephasings defined by 
\beq
Q_{j L} = e^{- i \alpha_j} Q_{j L}'
\label{qrephase}
\eeq
\beq
u_{j R} = e^{i \beta^{(u)}_j} u_{j R}'
\label{urephase}
\eeq
\beq
d_{j R} = e^{i \beta^{(d)}_j} d_{j R}'
\label{drephase}
\eeq
for $1 \le j \le N_G$. (The minus sign in (\ref{qrephase}) is included for 
technical convenience to avoid minus signs in later formulas.)  
In passing, we note that since the mass terms in eq. (\ref{massterm}) are
assumed to have arisen from $G_{SM}$-invariant Yukawa interactions, the Yukawa
matrices, as coefficients of $G_{SM}$-invariant operators, are obviously 
left invariant by $G_{SM}$ gauge transformations.  Also, note that since 
at the level of (\ref{massterm}) which defines the quark mass matrices, no 
explicit Higgs fields appear, we do not apply the rephasings of these Higgs
fields.  In any case, in particular models, such as the MSSM, the original
Higgs doublets $H_1$ and $H_2$ have already been rephased so as to yield real
vacuum expectation values $v_d$ and $v_u$.  In terms of the primed (rephased) 
fermion fields, the mass matrices, or equivalently the Yukawa matrices,
have elements
\beq 
Y_{jk}^{(f) \prime} = e^{i(\alpha_j + \beta^{(f)}_k)}Y_{jk}^{(f)}
\label{yfrephased}
\eeq
for $f=u,d$. 
Thus, if $Y^{(f)}$ has $N_f$ nonzero, and, in general, complex elements, then
the $N_f$ equations for making these elements real are
\beq
\alpha_j + \beta^{(f)}_k = -\arg(Y_{jk}^{(f)}) + \eta^{(f)}_{jk}\pi
\label{yfrephaseq}
\eeq
for $f=u,d$, where the set $\{jk\}$ runs over each of these nonzero elements, 
and $\eta^{(f)}_{jk} = 0$ or $1$.  The $\eta_{jk}$ term allows for the 
possibility of making the rephased element real and negative rather than
positive; this will not affect the counting of unremovable phases.
Let us define the vector of fermion field phases 
\beq
v = (\{\alpha_i \}, \{ \beta^{(u)}_i \}, \{\beta^{(d)}_i\})^T
\label{vvector}
\eeq
of dimension $3N_G$, where 
$\{\alpha_i\} \equiv \{\alpha_1,..., \alpha_{N_G}\}$, 
$\{\beta^{(f)}_i\} \equiv \{\beta^{(f)}_1,...,\beta^{(f)}_{N_G}\}$ for 
$f=u,d$, and 
\beq
w=(\{-\arg(Y^{(u)}_{jk})+\eta^{(u)}_{jk}\pi \}, \ \{-\arg(Y^{(d)}_{mn})+
\eta^{(d)}_{mn}\pi 
\})^T
\label{wvector}
\eeq
of dimension equal to the number of rephasing equations $N_{eq} = N_u +
N_d$, where the indices $jk$ and $mn$ take appropriate values.\footnote{In 
eq. (\ref{wvector}) we change notation from Ref. \cite{ks}, 
where an overall minus sign was absorbed in $T$, so that the analogue of eq. 
(\ref{wvector}) was 
\beq
w=(\{\arg(Y^{(u)})+\eta^{(u)}_{jk}\pi \}, \ \{\arg(Y^{(d)})+\eta^{(d)}_{jk}\pi 
\})^T
\label{otherw}
\eeq
(recall that since the arguments are defined mod $2\pi$ and
$\eta^{(f)}_{jk}=0$ or $1$, it follows that $\eta^{(f)}_{jk}\pi = 
-\eta^{(f)}_{jk}\pi$ (mod $2\pi$).)  The present definition avoids minus 
signs in our explicit listings of $T$ matrices below (e.g. eq.
(\ref{t1}) etc.).}
For explicit calculations, the ordering of the elements of the
vector $w$ is given by the sequence of nonzero elements $Y^{(u)}_{jk}$ for 
$jk=\{11,12,13,21,22,23,31,32,33\}$ followed by the nonzero elements 
$Y^{(d)}_{mn}$ with $mn=$ ranging over the same set, 
$\{11,12,13,21,22,23,31,32,33\}$.
We can then write (\ref{yfrephaseq}) for $f=u,d$ as 
\beq
T v = w
\label{teq}
\eeq
which defines the $N_{eq}$-row by $3N_{G}$-column matrix $T$.  Since a nonzero
element of $Y^{(f)}$, $f=u,d$ is, in general complex (because no symmetry
requires it to be real), there is an obvious general relation between the 
number of rephasing equations $N_{eq}$, its maximal value, 
$(N_{eq})_{max}=2N_G^2$, and the number of zero elements, $N_z$, in the Yukawa 
matrices:
\beq
N_{eq}=2N_G^2-N_z
\label{neqnz}
\eeq
In studying specific models, it is also helpful to introduce the notation
$N_{z,f}$ for the number of zeroes in each of the Yukawa matrices $Y^{(f)}$,
for $f=u$ and $f=d$.  Obviously, $N_z=N_{z,u}+N_{z,d}$.  If an element of 
$Y^{(f)}$, $f=u,d$ is zero at some energy scale $E$, this will not 
necessarily be true at a different energy scale $E'$, so that in specifying 
the numbers $N_{z,u}$ and $N_{z,d}$, as well as in specifying the
forms of the Yukawa matrices $Y^{(u)}$ and $Y^{(d)}$ themselves, one must
indicate the energy scale involved.  If at this energy scale, the gauge group
and the structure of the Yukawa couplings guarantee that the $Y^{(f)}$, 
$f=u,d$, are (complex) symmetric, then, of course, $Y^{(f)}_{jk}=0$ is 
equivalent to $Y^{(f)}_{kj}=0$; however, it will be convenient to count both in
our definition of $N_{z,f}$ and $N_z$. 

    Since $T$ is an $N_{eq}$-row by $3N_{G}$-column matrix, clearly 
\beq
rank(T) < \min \{ N_{eq}, \ 3N_G \}
\label{rt1}
\eeq
For the case (realized in all experimentally viable models which we have
studied) where $N_{eq} \ge 3N_G$, one has a more restrictive upper bound on
$rank(T)$: 
\beq
rank(T) \le 3N_G-1 , \ \quad \quad (for \ \ N_{eq} \ge 3N_G)
\label{rankt}
\eeq
This is proved by ruling out the only other possibility, i.e. $rank(T)=3N_G$. 
The reason that $rank(T)$ cannot have its apparently maximal value is that 
one overall rephasing has no effect 
on the Yukawa interaction, namely the U(1) generated by
(\ref{qrephase})-(\ref{drephase}) with $-\alpha_i =
\beta^{(u)}_j = \beta^{(d)}_k$ for all $i,j,k$.  For the relevant case $N_G=3$,
for all of the (realistic) models that we have studied, the upper bound 
(\ref{rankt}) is saturated, i.e. $rank(T)=8$.  This is not a general result; we
will exhibit a toy model (model 7 in section 5) where $rank(T)=7$, although 
that model is a degenerate case. 

   The main theorem~\cite{ks} is as follows:  \newline
{\it Theorem}: \newline
The number of unremovable phases $N_p$ in $Y^{(u)}$ and $Y^{(d)}$ is 
\beq
N_p = N_{eq}- rank(T)
\label{np}
\eeq
(which holds for arbitrary $N_G$).  \newline
\noindent {\it Proof:} \newline
First, denote $rank(T)=r_{_T}$.  Then one can select and delete 
$N_{eq}-r_{_T}$ rows from the matrix $T$ without reducing the rank of the 
resultant matrix.  This deletion means that we do not attempt to remove the 
phases from the corresponding elements of the $Y^{(f)}$, $f=u,d$.  We perform 
this reduction. For the remaining $r_{_T}$ equations, we move a subset of 
$3N_G-r_{_T}$ phases in $v$ 
to the right-hand side of (\ref{yfrephaseq}), thus including them in a 
redefined $\bar w$.  This yields a set of $r_{_T}$ linear equations in 
$r_{_T}$ unknown phases, denoted $\bar v$. We write this as $\bar T \bar v = 
\bar w$.  Since by construction $rank(\bar T)=r_{_T}$, it follows that 
$\bar T$ is invertible, so that one can now solve for the $r_{_T}$ fermion 
rephasings in $\bar v$ which render $r_{_T}$ of the $N_{eq}$ complex elements 
real: $\bar v = \bar T^{-1}\bar w$. Hence there are $N_{eq}-r_{_T}$ remaining 
phases in the $Y^{(f)}$, as claimed. $\Box$.  \newline
The operations involved in this proof and the construction of $\bar T$ will 
be explicitly illustrated for model 1 in section 5.

  Some comments are in order.  First, we note that 
in order for these phases to be physically meaningful, it is, of course,
necessary that they be unchanged under the full set of rephasings of fermion
fields.  In fact, as we will show, there is a one-to-one correspondence
between each such unremovable phase and an independent phase of a certain
product of
elements of mass matrices which is invariant under fermion field rephasings.
(In special cases a model may have an unremovable, invariant phase which,
because of a particular symmetry, is zero or $\pi$.) 
Second, as is clear from our proof, in general,
the result  (\ref{np}) does not depend on whether or not
$Y^{(f)}_{jk} = Y^{(f)}_{kj}$ initially.  Hence making $Y^{(f)}$ (complex)
symmetric for either $f=u$ or $f=d$ does not, in general, result in any 
reduction in $N_p$, since in terms of the $G_{SM}$ theory, nothing guarantees 
this symmetry, and the rephasing is carried out in terms of $G_{SM}$ fields.  
Third, if one of the unremovable phases is put in a given off-diagonal 
$Y^{(f)}_{jk}$, one may wish to modify the $kj'th$ equation to read 
\beq
\alpha_k + \beta^{(f)}_j = -\arg(Y_{kj}^{(f)})-\arg(Y_{jk}^{(f)})
\label{yfhermitian}
\eeq
For example, in a model where $|Y^{(f)}_{jk}|=|Y^{(f)}_{kj}|$, this would 
yield $Y_{jk}^{(f)*} = Y_{kj}^{(f)}$ for this pair $jk$.  The modification in
(\ref{yfhermitian}) has no effect on the counting of phases. 

    Using our theorem, we may calculate the maximal number of unremovable
phases $N_p$ in the Yukawa matrices as a function of the number of fermion 
generations, $N_G$.  For this purpose, we consider the case in which these
matrices have all nonzero arbitrary (and, hence in general, complex) elements. 
Then $N_{eq}$ takes on its maximal value, 
\beq
(N_{eq})_{max}=2N_G^2
\label{neqmax}
\eeq
Since the elements of the $Y^{(f)}$ are arbitrary, it follows that they satisfy
no special relations, so that the $T$ matrix has maximal rank.  By our result
(\ref{rankt}), this is $rank(T)_{max}=3N_G-1$.  Using our theorem (\ref{np}), 
we thus find that the maximal number of unremovable phases is
\beq
(N_p)_{max}=(N_G-1)(2N_G-1)
\label{npmax}
\eeq
That this case yields the maximal number of unremovable phases is clear, since
we can obtain any other form by making some entries of the Yukawa matrices 
zero, and this can only decrease the number of unremovable phases. 
For $N_G=1,2$ and 3, $(N_p)_{max}$ takes the values 0, 3, and 10, 
respectively.  In current phenomenological models of quark mass matrices, 
one tries to make as many entries in the Yukawa matrices as possible zero to
minimize the number of parameters and increase predictiveness, so that in the
physical $N_G=3$ case, we will find that $N_p$ is typically just 2 or 3.

\section{Rephasing Invariants and Theorems on Locations of Phases}
\label{invariants}

   A fundamental question concerns which elements of $Y^{(u)}$ and $Y^{(d)}$
can be made real by fermion rephasings.  This is connected with the issue of 
which rows are to be removed from $T$ to obtain $\bar T$, i.e. 
which nonzero elements of the $Y^{(u)}$ and $Y^{(d)}$ are left complex.  In
Ref. \cite{ks} several theorems were given which provide a general answer to 
these questions.  We review these theorems here and give the details of the
proofs.  As in the previous section, for generality, we shall indicate which 
results apply for an arbitrary number of fermion generations, $N_G$.  In the
next section we shall also give a number of new results for arbitrary $N_G$. 
The general method is to construct all independent complex products of 
elements of the $Y^{(f)}$, $f=u,d$, having the property that these products 
are invariant under the rephasings (\ref{qrephase})-(\ref{drephase}).  Here we
use the term ``invariant products'' to mean invariant under the rephasings 
(\ref{qrephase})-(\ref{drephase}).  
These products must involve an even number of such elements, since for each 
index on a $Y$, there must be a corresponding index on an $Y^*$ in order to 
form a rephasing-invariant quantity.  Since in general, by construction, 
these invariant products are complex, i.e. have arguments $\ne 0,\pi$, each 
one implies a constraint which is that the
set of $2n$ elements which comprise it cannot be made simultaneously real by
the rephasings (\ref{qrephase})-(\ref{drephase}) of the quark fields. 
Here, by ``complex invariant products'', 
we mean products which are, in general, 
complex for arbitrary $Y^{(f)}$; in special cases, a symmetry of a given 
model may render some of these complex invariants real.  It is easily seen that
the only rephasing-invariant products of two elements of the $Y^{(f)}$ must 
be of the form $|Y^{(f)}_{jk}|^2$, i.e., there are no complex invariant 
products of order 2.  We define an irreducible complex invariant to be one
which cannot be factorized purely 
into products of lower-order complex invariants.
Given that there are no quadratic complex invariants, it follows that all
complex invariants of order 4 and 6 are irreducible.  As we will prove below,
for the physical case of $N_G=3$ generations of fermions, the phase constraints
are totally determined by the complex invariants of order 4 and 6, so in the
analysis of phenomenological models, one automatically deals only with
irreducible complex invariants.\footnote{Reducible complex invariants only 
occur in the hypothetical case $N_G \ge 4$ and then first occur at order 8, 
as products of different complex quartic invariants.  Since reducible 
invariants do not yield any new phase constraints, it suffices to consider only
irreducible complex invariants.  Hence in all of our discussion below, in
particular, the section on arbitrary $N_G$, we shall take ``complex 
invariants'' to mean ``irreducible complex invariants''.}  We define a set of
independent (irreducible) complex invariants to be a set of complex
(irreducible) invariants with the property that no invariant in the set is
equal to (i) the complex conjugate of another invariant in the set or (ii)
another element in the set with its indices permuted.  This does not imply that
the arguments of a set of independent (irreducible) complex invariants are 
linearly independent.  
Indeed, we will show examples below of sets of independent 
complex invariants of the same order whose arguments are linearly dependent,
and also examples of higher-order complex invariants whose arguments can be
expressed as linear combinations of arguments of lower-order complex
invariants.  We define $N_{ia}$ to
denote the number of linearly independent arguments ($\rightarrow ia$) among
the $N_{inv}$ independent complex invariants.  Further, we define $N_{inv}$ 
to be the total number of independent (irreducible) complex invariants in a 
given model.  It is useful to define $N_{inv,2n}$ to denote the number of
independent complex invariants of order $2n$.

     We thus construct a set of rephasing-invariant products 
depending on the up and down quark sectors individually:
\beq
P^{(f)}_{2n;j_1 k_1,...j_n k_n;\sigma_L} = 
\prod_{a=1}^{n}Y^{(f)}_{j_a k_a}Y^{(f)*}_{\sigma_L(j_a) k_a}
\label{pgeneral}
\eeq
where $f=u,d$, and $\sigma_L$ is an element of the permutation group $S_n$. 
(This set of invariants is clearly the same as the sets defined by 
$\prod_{a=1}^{n}Y^{(f)}_{j_a k_a}Y^{(f)*}_{j_a \sigma_R(k_a)}$ and 
$\prod_{a=1}^{n}Y^{(f)}_{j_a k_a}Y^{(f)*}_{\sigma_L(j_a) \sigma_R(k_a)}$, where
$\sigma_R \in S_n$; with no loss of generality, we write it in the form of eq.
(\ref{pgeneral}).)  The first theorem is: The products in (\ref{pgeneral}) are
invariant under the rephasings of fermion fields
(\ref{qrephase})-(\ref{drephase}).  To prove this, we observe that from eq. 
(\ref{yfrephased}), it follows that under these rephasings, 
\begin{eqnarray}
\prod_{a=1}^n Y^{(f)}_{j_a k_a}Y^{(f)*}_{\sigma_L(j_a) k_a} & \to &
\prod_{a=1}^n [e^{i(\alpha_{j_a}+\beta^{(f)}_{k_a})}Y^{(f)}_{j_a k_a}]
[e^{-i(\alpha_{\sigma_L(j_a)}+\beta^{(f)}_{k_a})}Y^{(f)*}_{\sigma_L(j_a) k_a}] 
\nonumber \\
& = & \Bigl ( e^{i\sum_{b=1}^n (\alpha_{j_b}-\alpha_{\sigma_L(j_b)})} 
\Bigr ) 
\prod_{a=1}^n Y^{(f)}_{j_a k_a} Y^{(f)*}_{\sigma_L(j_a) k_a} \nonumber \\
& = & \prod_{a=1}^n Y^{(f)}_{j_a k_a} Y^{(f)*}_{\sigma_L(j_a) k_a} 
\label{pinvariance}
\end{eqnarray}
where $\sum_{b=1}^n (\alpha_{j_b} -\alpha_{\sigma_L(j_b)})=0$ because 
the permutations of $S_n$ are an automorphism of the set $(1,2,...,n)$.  Hence,
the products $P^{(f)}_{2n;j_1 k_1,...j_n k_n;\sigma_L}$ are invariant under
rephasings of the quark fields, as claimed. $\Box$.

  Secondly, we construct a set of invariants connecting the up and down quark
sectors: 
\beq
Q^{(s,t)}_{2n;\{j\},\{k\},\{m\};\sigma_L,\sigma_u;\sigma_d} = 
(\prod_{a=1}^{s}Y^{(u)}_{j_a k_a})
(\prod_{b=1}^{t}Y^{(d)}_{j_{s+b} \ m_b})
(\prod_{c=1}^{s}Y^{(u)*}_{\sigma_L(j_c)\sigma_u(k_c)})
(\prod_{e=1}^{t}Y^{(d)*}_{\sigma_L(j_{s+e})\sigma_d(m_e)})
\label{qgeneral}
\eeq
where $s, t \ge 1$, $s+t=n$, $\sigma_L \in S_n$, $\sigma_u \in S_s$, and 
$\sigma_t \in S_t$.  The proof that these are invariant proceeds in a manner
similar to that just given above; hence we do not give it explicitly.  
At quartic order, $2n=4$, $\sigma_L \in S_2$ in eq. (\ref{pgeneral}).  
We use the
standard notation $(^{12}_{12}) \equiv 1$ and $(^{12}_{21}) \equiv (12) \equiv
\tau$ (transposition) for the elements of $S_2$.  For $\sigma_L=\tau$, we 
obtain the complex invariants 
\beqs
P^{(f)}_{j_1 k_1, j_2 k_2} & \equiv &
 P^{(f)}_{4;j_1 k_1, j_2 k_2;\sigma_L=\tau} \nonumber \\
                                            \nonumber \\
& = & Y^{(f)}_{j_1 k_1}Y^{(f)}_{j_2 k_2}Y^{(f)*}_{j_2 k_1}Y^{(f)*}_{j_1 k_2}
\label{p}
\eeqs
for $f=u,d$.  (For the other choice, $\sigma_L=1$, the resultant invariant 
product, $P^{(f)}_{4;j_1 k_1, j_2 k_2;\sigma_L=1}=
|Y^{(f)}_{j_1 k_1}|^2|Y^{(f)}_{j_2 k_2}|^2$ is real and hence yields no 
constraint on phases.)
At this quartic order, there is only one $Q$-type complex invariant; 
this has $s=t=1$, $\sigma_L=\tau$, so that in the general notation of
(\ref{qgeneral}), it is 
$Q^{(1,1)}_{4;\{j_1,j_2\},\{k_1\},\{m_1\};\sigma_L=\tau,\sigma_u=1,
\sigma_d=1}$.
For brevity of notation, we denote it simply as 
\beqs
Q_{j_1 k_1,j_2 m_1} & \equiv & 
Q^{(1,1)}_{4;\{j_1,j_2\},\{k_1\},\{m_1\};\sigma_L=\tau,\sigma_u=1}
\nonumber \\ 
\nonumber \\
& = & Y^{(u)}_{j_1 k_1}Y^{(d)}_{j_2 m_1}Y^{(u)*}_{j_2 k_1}Y^{(d)*}_{j_1 m_1}
\label{q}
\eeqs
(As with the quartic $P$ invariant, the other choice for the permutation, 
$\sigma_L=1$, yields the product $|Y^{(u)}_{j_1 k_1}Y^{(d)}_{j_2 m_1}|^2$, 
which is real.)
Note the index symmetry 
\beq
P^{(f)}_{j_1 k_1, j_2 k_2}=P^{(f)}_{j_2 k_2, j_1 k_1}
\label{psym}
\eeq
and the complex conjugation relations
\beq
P^{(f)}_{j_1 k_1 j_2 k_2}=P^{(f)*}_{j_1 k_2, j_2 k_1}
\label{pconj}
\eeq
and
\beq
Q_{j_1 k_1, j_2 m_1}=Q_{j_2 k_1, j_1 m_1}^*
\label{qconj}
\eeq

    At order $2n=6$, we find one independent $P$-type complex invariant for 
each quark sector $f=u,d$, and two independent $Q$-type complex invariants;
these were given in Ref. \cite{ks}.  Here we present the details of the 
analysis. For a given $f$, we construct the invariants corresponding to each 
of the six permutations $\sigma_L \in S_3$.  In standard notation, these 
permutations are listed as 
\beq
S_3 = \{ (^{123}_{123}) \equiv 1, \ (^{123}_{213}) \equiv (12), \ 
(^{123}_{132}) \equiv (23), \ (^{123}_{321}) \equiv (13), \ 
(^{123}_{231}) \equiv (231), \ (^{123}_{312}) \equiv (312)\}
\label{s3}
\eeq
Clearly for $\sigma_L=1$, 
$P^{(f)}_{6;j_1 k_1, j_2 k_2, j_3 k_3;\sigma_L=1} = |Y^{(f)}_{j_1 k_1}
Y^{(f)}_{j_2 k_2}Y^{(f)}_{j_3 k_3}|^2$, which is real and hence implies no
constraint on phases.  For the three transpositions $\sigma_L = (12), (23),
(13)$, the resultant product reduces to a real factor times a quartic $P$ 
invariant; for example, $P^{(f)}_{6;j_1 k_1, j_2 k_2, j_3 k_3;\sigma_L=(12)} = 
|Y^{(f)}_{j_3 k_3}|^2 P^{(f)}_{4;j_1 k_1, j_2 k_2}$, and so forth for the
others. The last two elements of $S_3$ yield genuine 6'th order complex
invariants: 
\beq
P^{(f)}_{6;j_1 k_1, j_2 k_2, j_3 k_3;\sigma_L=(231)} = 
Y^{(f)}_{j_1 k_1}Y^{(f)}_{j_2 k_2}Y^{(f)}_{j_3 k_3}
Y^{(f)*}_{j_2 k_1}Y^{(f)*}_{j_3 k_2}Y^{(f)*}_{j_1 k_3}
\label{p6a}
\eeq
and 
\beq
P^{(f)}_{6;j_1 k_1, j_2 k_2, j_3 k_3;\sigma_L=(312)} = 
Y^{(f)}_{j_1 k_1}Y^{(f)}_{j_2 k_2}Y^{(f)}_{j_3 k_3}
Y^{(f)*}_{j_3 k_1}Y^{(f)*}_{j_1 k_2}Y^{(f)*}_{j_2 k_3}
\label{p6b}
\eeq
However, only one of these is independent, since they are related according to
\beq
P^{(f)}_{6;j_1 k_1, j_2 k_2, j_3 k_3;\sigma_L=(231)} = 
P^{(f)}_{6;j_3 k_1, j_1 k_2, j_2 k_3;\sigma_L=(312)}
\label{p6abrel}
\eeq
This explicit calculation proves our assertion above, that there is one 
(independent) 6'th order $P$-type complex invariant.  We may take this to be 
the one in eq. (\ref{p6b}), and we denote it in a simple notation as 
\beq 
P^{(f)}_{j_1 k_1, j_2 k_2, j_3 k_3} \equiv
P^{(f)}_{6;j_1 k_1, j_2 k_2, j_3 k_3;\sigma_L=(312)}
\label{p6}
\eeq

To find the 6'th order complex $Q$-type invariants, we consider the case
$(s,t)=(2,1)$ in eq. (\ref{qgeneral}); the other case $(s,t)=(1,2)$ can then be
obtained easily by the replacement $Y^{(u)} \leftrightarrow Y^{(d)}$ in the
resultant products. Since $\sigma_L \in S_3$, $\sigma_u \in S_2$, and
$\sigma_d = 1$, it follows that there are $3! \times 2!=12$ invariants of 
this type.
Each of these can be labelled by $\sigma_L \otimes \sigma_u$.  Clearly the $1
\otimes 1$ and $(12) \otimes (12)$ products are real.  Further, we find 
that the products $1 \otimes (12)$, and 
$(12) \otimes (12)$ give real factors times quartic $P$-type invariants, while
$(23) \otimes 1$, $(13) \otimes 1$, $(231) \otimes (12)$, and
$(312) \otimes (12)$ give real factors times quartic $Q$-type
invariants.  The remaining four choices for the permutations give genuine
complex 6'th order invariants.  To discuss these, we shall introduce a
convenient notation, 
\beq
Q^{(2,1)}_{6;j_1 k_1, j_2 k_2, j_3 m_1;\sigma_L,\sigma_u} \equiv
Q^{(2,1)}_{6;\{j_1,j_2,j_3\},\{k_1,k_2\},\{m_1\};\sigma_L,\sigma_u 
(\sigma_d=1)}
\label{qequiv}
\eeq
The permutations 
$(23) \otimes (12)$ and $(312) \otimes 1$ yield the same product, 
\beq
Q^{(2,1)}_{6;j_1 k_1, j_2 k_2, j_3 m_1;\sigma_L=(23),\sigma_u=(12)} = 
Y^{(u)}_{j_1 k_1}Y^{(u)}_{j_2 k_2}Y^{(d)}_{j_3 m_1}
Y^{(u)*}_{j_1 k_2}Y^{(u)*}_{j_3 k_1}Y^{(d)*}_{j_2 m_1}
\label{q6a}
\eeq
while the choices $(13) \otimes (12)$ and $(231) \otimes 1$ yield the
same product, 
\beq
Q^{(2,1)}_{6;j_1 k_1,j_2 k_2,j_3 m_1;\sigma_L=(13),\sigma_u=(12)} = 
Y^{(u)}_{j_1 k_1}Y^{(u)}_{j_2 k_2}Y^{(d)}_{j_3 m_1}
Y^{(u)*}_{j_3 k_2}Y^{(u)*}_{j_2 k_1}Y^{(d)*}_{j_1 m_1}
\label{q6b}
\eeq
However, (\ref{q6a}) and (\ref{q6b}) are not independent, since 
\beq
Q^{(2,1)}_{6;j_1 k_1, j_2 k_2, j_3 m_1;\sigma_L=(13),\sigma_u=(12)} = 
Q^{(2,1)}_{6;j_3 k_1, j_1 k_2, j_2 m_1;\sigma_L=(23),\sigma_u=(12)}
\label{q6abrel}
\eeq
Hence, as stated above, there are only two independent complex 6'th order
invariants, which may be taken to be (\ref{q6a}) and its mirror pair under
$Y^{(u)} \leftrightarrow Y^{(d)}$. For 
brevity of notation, we shall denote these simply as 
\beq
Q^{(fff')}_{j_1 k_1, j_2 k_2, j_3 m_1} = 
Y^{(f)}_{j_1 k_1}Y^{(f)}_{j_2 k_2}Y^{(f')}_{j_3 m_1}
Y^{(f)*}_{j_1 k_2}Y^{(f)*}_{j_3 k_1}Y^{(f')*}_{j_2 m_1}
\label{q6}
\eeq
where $(fff')=(uud)$ or $(ddu)$.  Note the relations 
\beq
P^{(f)}_{6;j_1 k_1,j_2 k_2, j_3 k_3} = P^{(f)*}_{j_3 k_2, j_2 k_1, j_1 k_3}
\label{p6conj}
\eeq
and
\beq
Q^{(fff')}_{6;j_1 k_1, j_2 k_2, j_3 m_1} = 
Q^{(fff')*}_{6;j_1 k_2,j_3 k_1, j_2 m_1}
\label{q6conj}
\eeq
So far, all of the results in this section hold for arbitrary $N_G$.  For the
physical case $N_G=3$, the order 4 and order 6 invariants given above form a 
complete set, since, as was noted in Ref. \cite{ks} (for $N_G=3$) there are 
are no new constraints from any invariant of order $\ge 8$ (see below).  
We shall sometimes refer to $P$ and $Q$ invariants of order 4 and order 6 
generically via the respective symbols $P_4$, $Q_4$, $P_6$, and $Q_6$. 
  
    We next review our theorems \cite{ks} and present the proofs. 
For a given model, construct the maximal set of $N_{ia}$ independent complex 
invariants of lowest order(s), whose arguments (phases) are linearly 
independent. Then we have the \newline
{\it Theorem}: \newline
(a) each of these invariants implies a constraint that the elements contained 
within it cannot, in general, all be made 
simultaneously real; (b) this constitutes the complete set of
constraints on which elements of $Y^{(u)}$ and $Y^{(d)}$ can be made 
simultaneously real; and hence (c) $N_p = N_{ia}$. These results hold for
arbitrary $N_G$.  Further, (d) for the physical case $N_G=3$, the complex
invariants of order 4 and 6 form a complete set, i.e., the argument of any 
complex invariant of order $\ge 8$ can be expressed in terms of the arguments
of complex invariants of orders 4 and 6, and hence these higher order 
complex invariants do not yield any new phase constraints.  Hence also, for
the physical case $N_G=3$, the total number of independent (irreducible) 
complex invariants is given by
\beq
N_{inv}=N_{inv,4}+N_{inv,6}, \ \quad for \ \ N_G=3
\label{ninvng3}
\eeq

(A discussion of 
complex invariants in the case of arbitrary $N_G$ is given in section
\ref{generalng}.) 

{\it Proof:} The proof of (a) is essentially obvious: since by 
construction, the argument of each complex invariant is 
invariant under the rephasings of the fermion fields 
(\ref{qrephase})-(\ref{drephase}) and since it is not, in general, equal 
to 0 or $\pi$, the corresponding invariant cannot in general be made real.  It
follows, {\it a \ fortiori}, that it is not, in general, possible to make the
Yukawa matrix elements which comprise this complex invariant product all
simultaneously real.  Given that this is a maximal set, the assertions in parts
(b) and (c) of the theorem follow immediately.  The proof of (d) is more
involved. Let $X_{2n}$ be some non-zero invariant of type $P$ or $Q$.
Further, let $Y_{ij}^{(f)}$, $f = u$ or $d$, be one of the elements in 
$X_{2n}$. Then it is clear that in order for $X_{2n}$ to be an invariant,
it must also
contain the product of $Y_{ik}^{(f')*} \: Y_{lj}^{(f)*}$ for some $k,l=1,2,3$,
where $f'$ may be equal to, or different from, $f$.  Hence we can write 
\beq
X_{2n}= Y_{ij}^{(f)} Y_{lj}^{(f)*} Y_{ik}^{(f')*} X'
\label{x2ndecomp}
\eeq
Now we consider the
following two cases. (i) For some values of $k$ and $l$, the Yukawa matrix 
element $Y_{lk}^{(f')}$ is non-zero.  Then 
\begin{eqnarray}
\arg(X_{2n})= \arg(Y_{lk}^{(f')*}Y_{lk}^{(f')} \: X_{2n}) =  \ \ \ \ \ \ \ \ 
\nonumber \\
\arg(
Y_{lk}^{(f')*} \: (Y_{lk}^{(f')} Y_{ij}^{(f)} Y_{ik}^{(f')*} Y_{lj}^{(f)*} )
\: X' )= arg ( \tilde{X_{4}} \: \tilde{X}_{2n-2} )  
\label{insertion}
\end{eqnarray}
where 
\beq
\tilde{X_{4}} = Y_{lk}^{(f')} Y_{ij}^{(f)} Y_{ik}^{(f')*} Y_{lj}^{(f)*}
\label{x4tilde}
\eeq
and 
\beq
\tilde X_{2n-2} = Y^{(f')}_{lk}X'
\label{x2nm2}
\eeq
Hence, 
\beq
\arg(X_{2n}) = \arg(\tilde X_4) + \arg(\tilde X_{2n-2})
\label{argreduce}
\eeq
Observe that eq. (\ref{x4tilde}) subsumes both the case where $\tilde X_4$ is 
of $P$ type and the case where it is of $Q$ type.  Given the definition
(\ref{x2ndecomp}), it is easily seen that (\ref{x2nm2}) is an invariant. 
Therefore, in this case the argument of the order $2n$ invariant $X_{2n}$ can 
be expressed as a sum of the arguments of an order 4 invariant $\tilde X_4$ 
and an order $(2n-2)$ invariant
$\tilde X_{2n-2}$, so that $\arg(X_{2n})$ does not yield a new constraint. 
By induction, every invariant of order greater than 4 which
satisfies the conditions of (i) is reducible in this manner. 

   It remains to consider the other case, that (ii) for all values of 
$i,j,k$ and $l$, $Y_{lk}^{(f')}=0$.  In this case, the invariant $X_{2n}$
cannot be reduced by means of the above procedure.  Then it is also true that
since $Y_{lk}^{(f')}$ vanishes, it cannot be one of the elements in 
the $X_{2n}$, because the latter product is nonzero by assumption.
Consequently, the invariant $X_{2n}$, which contains the product 
$ Y_{ij}^{(f)} Y_{lj}^{(f)*} Y_{ik}^{(f')*} X'$, cannot contain the element 
$Y_{lk}^{(f')}=0$ at the same time.  Because each invariant contains either 
no elements, or at least two elements, from each
column of the Yukawa mass matrices (in the latter case, denote the two as 
$Y_{pr}^{(f)}$ and $Y_{qr}^{(f)*}$), it follows that the $X_{2n}$ invariant 
in this case  
must be constructed out of blocks $B_{pqr}^{(f)}=(Y_{pr}^{(f)} Y_{qr}^{(f)*})$ 
such that the pairs of the first indices, $(p,q)$, of the two such blocks 
do not coincide:
\beq
X_{2n}= \prod_{i} B_{p_{i} \: q_{i} \: r}^{(f_{i})}, \ where \ 
(p_{i}, q_{i}) \neq (p_{j}, q_{j})  \ \ if \ \ i \neq j
\eeq
Since $p$ and $q$ can take only 3 values each, there are 
$(^{3}_{2})=3 $ possibilities to assign these indices (where
$(^m_n)=m!/(n!(m-n)!)$ denotes the binomial coefficient). 
This means that the order of an invariant which is not decomposable by these 
means cannot be greater than 3 times the order of $B$, i.e., 6.  We can 
exhibit examples (see section 5) for 
 which this bound is saturated, i.e. the order
of the lowest-order nonvanishing complex invariant is 6.  This thus proves 
(d). $\Box$   \newline  For most models the lowest-order nonvanishing complex
invariants are of order 4, but, as mentioned, we will in section 5 exhibit 
some exceptions for which the lowest-order nonvanishing complex invariants 
are of order 6.

{\it Corollary 1:} \newline
If in a given model there are as at least as many independent quartic 
complex invariants with independent arguments as there are unremovable phases,
then the arguments of not only order 8 and higher, but also of 6'th order 
complex invariants are expressible in terms of those of the quartic invariants
and hence yield no new phase constraints. \newline
{\it Proof:} \newline
This is obvious, since the premise of the corollary means that since all of the
$N_p=N_{ia}$ unremovable phases have already been accounted for by quartic
complex invariants, there cannot be any further phase constraints from any 
higher order complex invariants, and hence the arguments of all such higher
order complex invariants must be expressible in terms of those of the $N_{ia}$
quartic complex invariants with independent arguments.  $\Box$  \newline
For most models, the lowest-order complex invariants are of 
order 4, but, as mentioned, we will exhibit two exceptions in section 5, where
the only complex invariants are of order 6.  
   
{\it Corollary 2:} \newline
If in a given model all of the elements of the $Y^{(f)}$,
$f=u,d$ are nonzero arbitrary (complex) quantities, then the arguments of the
order 6, as well as the higher order, complex invariants are expressible in
terms of those of the quartic invariants.  \newline
{\it Proof} \newline
This is clear, since given any order 6 complex invariant, we can always apply
the reduction method of eqs. (\ref{insertion})-(\ref{argreduce}) to express its
argument in terms of that of the arguments of two quartic complex invariants. 
$\Box$ \newline
(Recall that in this case such a model would also have the maximal 
number of unremovable phases, (\ref{npmax}).)

    We have developed a graphical representation for the complex invariants 
which we have found useful.  This is described next.  The Yukawa matrix 
$Y^{(f)}$ is a $N_G \times N_G$ matrix whose rows and columns are labelled
respectively by the first and second indices, $j$ and $k$ on the elements
$Y^{(f)}_{jk}$.  From the defining equation, (\ref{p}), we can represent 
the quartic invariant $P^{(f)}_{j_1 k_1, j_2 k_2}=
Y^{(f)}_{j_1 k_1}Y^{(f)}_{j_2 k_2}Y^{(f)*}_{j_2 k_1}Y^{(f)*}_{j_1 k_2}$ as
follows.  To the factor $Y^{(f)}_{j_1 k_1}Y^{(f)*}_{j_2 k_1}$ we associate an
oriented line segment (which can be visualized as a vertical arrow 
$\uparrow$ or $\downarrow$ connecting the two indicated elements at the
intersections of the rows $j_1$ and $j_2$ with the column $k_1$.  Let the 
head of the arrow or equivalently the $\odot$ of the dipole 
be placed on the unconjugated element, and the tail of the arrow or
equivalently the $\otimes$ of the dipole be placed on the conjugated element.
For the figures, it is actually convenient to use a notation involving vertical
dipoles $^{\odot}_{\otimes} = \uparrow$ and $^{\otimes}_{\odot} = \downarrow$,
where, as indicated, the $\odot$ of the dipole corresponds to the head of the
arrow (unconjugated element), while the $\otimes$ of the dipole corresponds to
the tail of the arrow (conjugated element). To the factor 
$Y^{(f)}_{j_2 k_2}Y^{(f)*}_{j_1 k_2}$ we associate a similar oriented vertical 
line segment connecting the elements at the intersections of the $j_2$ and 
$j_1$ rows with the column  $k_2$.  Note that the head of this arrow points 
in the opposite direction from the first arrow, or equivalently, the dipole is
oppositely oriented.  These two opposite arrows link the same rows;
consequently, they form a rectangle.  Hence, we may associate with each complex
quartic $P^{(f)}$ invariant a (possibly square) rectangle in the matrix 
$Y^{(f)}$.  As an example, our graphical representation of the quartic 
invariant $P^{(f)}_{22,33}$ is illustrated in Fig. 1(a).

    Similarly, we can represent the quartic invariant $Q_{j_1 k_1,j_2 m_1}=
Y^{(u)}_{j_1 k_1}Y^{(d)}_{j_2 m_1}Y^{(u)*}_{j_2 k_1}Y^{(d)*}_{j_1 m_1}$ as
follows.  Imagine the $Y^{(u)}$ matrix to lie next to the $Y^{(d)}$ matrix on a
plane.  To the factor 
$Y^{(u)}_{j_1 k_1}Y^{(u)*}_{j_2 k_1}$ we associate an oriented line
segment linking the elements at the intersections of the $j_1$ and $j_2$ rows 
with the $k_1$'th column in the $Y^{(u)}$ matrix.  As before, the head of the 
arrow is assigned to the unconjugated element.  To the other factor 
$Y^{(d)}_{j_2 m_1}Y^{(d)*}_{j_1 m_1}$ we associate an arrow linking the
elements at the intersections of the $j_2$ and $j_1$ rows with the $m_1$ column
in the $Y^{(d)}$ matrix.  As with $P$, the two arrows making up the $Q$ point
in opposite directions.  We may thus associate with each complex quartic $Q$ 
invariant a rectangle linking the arrow in the $Y^{(u)}$ matrix with the
oppositely oriented arrow in the $Y^{(d)}$ matrix.  As examples, the quartic
invariants $Q_{12,22}$ and $Q_{23,32}$ are illustrated in Fig. 1(b,c). 
The operation of complex
conjugation of a given complex quartic $P$ or $Q$ invariant reverses the 
directions of the arrows in each of the line segments.  Since the complex
conjugate invariants give the same phase constraints, the corresponding graphs
with reversed arrows may be considered to be equivalent to the graphs already
considered.  Hence for purposes of enumerating invariants, one may suppress the
degree of freedom associated with an overall reversal of all arrows (given that
the pairs of columns forming each graph always have oppositely directed
arrows), and thus deal with just the rectangles.  This is symbolically
indicated in Fig. 2 using the equivalent dipole notation. 

   Higher order invariants may be represented in a similar manner.  A 6'th
order $P$ complex invariant in the sector $f$ ($=u$ or $d$) is 
represented as three vertical arrows in three columns of $Y^{(f)}$ with the 
property that the head of each is matched by the tail of some other arrow.  A 
6'th order $Q^{(uud)}$ complex invariant is represented by two oppositely
directed  vertical arrows in $Y^{(u)}$ together with one vertical arrow in
$Y^{(d)}$, again with the property that the head of each arrow is matched by
the tail of some other arrow.  Finally, a 6'th order $Q^{(ddu)}$ invariant is
represented as for $Q^{(uud)}$ but with the interchange of $Y^{(u)}$ and
$Y^{(d)}$.  As examples, we show $Q^{(ddu)}_{32,23,12}$ and $P_{11,23,32}$ in
Figs. Figs. 3(a) and 3(b).

   Since the full set of $N_{inv}$ independent complex invariants 
will have arguments which are not, in general, linearly independent, it follows
that 
\beq 
N_{inv} \ge N_{ia}
\label{ninvia}
\eeq
It may also happen that, e.g. for order $2n=4$, the number of independent 
quartic complex invariants, $N_{inv,4}$ is greater than $N_{ia}$. 
For each complex invariant of a given order, $X$, 
\beq
\arg(X) = \sum_{f=u,d}\sum_{j,k}c^{(f)}_{j,k} \arg(Y^{(f)}_{jk})
\label{argx}
\eeq
where the sum is over the $N_{eq}$ complex elements of $Y^{(u)}$ and 
$Y^{(d)}$.  Including all orders of invariants, these equations can be 
written as 
\beq
\xi = Z w
\label{zeq}
\eeq
where $\xi$ is the $N_{inv}$-dimensional vector 
\beq
\xi = (\arg(X_1),...,\arg(X_{N_{inv}}))^T
\label{xidef}
\eeq
$Z$ is an $N_{inv}$-row by $N_{eq}$-column matrix, and it is implicitly
understood here that all of the $\eta^{(f)}_{jk}$ in eq. (\ref{wvector}) for 
$w$ are taken to be zero.  Often, some nonzero elements of $Y^{(u)}$ and 
$Y^{(d)}$ do not occur in any complex invariants and hence do not occur on the
right-hand side of (\ref{argx}).  These elements may be rephased freely. 
The $Z$ matrix as defined in eq. (\ref{zeq})
has columns of zeroes corresponding to each of these elements.  For simplicity,
one may thus define a reduced vector $w_r$ whose dimension is equal to the
number of elements of $Y^{(f)}$, $f=u,d$, which occur in complex invariants, 
and a corresponding reduced matrix $Z_r$ defined by
\beq
\xi = Z_r w_r
\label{zreq}
\eeq
Clearly, by removing columns of zeroes we do not reduce the rank, so
\beq
rank(Z_r)=rank(Z)
\label{zzrrank}
\eeq
Then the number of independent arguments among the complex invariants is 
given by the rank of $Z$:
\beq
N_{ia}=rank(Z)
\label{rankz}
\eeq
The eqs. (\ref{zeq}) and (\ref{zreq}) subsume the complex invariants of 
all orders (i.e. for the physical case $N_G=3$, orders 4 and 6).  One can also
apply this method to determine the linearly independent invariants of a given
order, $2n$.  For this purpose, we define a vector of complex invariants of
this order:
\beq
\xi_{2n} = (\arg(X_{2n;1}),...,\arg(X_{N_{inv,2n}}) )^T
\label{xi2n}
\eeq
We then have
\beq
\xi_{2n} = Z_{2n} w
\label{z2neq}
\eeq
As before, one can define a reduced vector $w_{r,2n}$ composed of the elements
of $Y^{(u)}$ and $Y^{(d)}$ which occur in the complex invariants of order $2n$,
and a corresponding reduced matrix $Z_{r,2n}$ satisfying
\beq
\xi_{2n} = Z_{r,2n} w_{r,2n}
\label{z2nreq}
\eeq
Then the number of linearly independent arguments among the complex invariants
of order $2n$ is given by $rank(Z_{2n})=rank(Z_{r,2n})$.

\section{Further Results for General $N_G$}
\label{generalng}

    In this section we give a number of new results on complex invariants and
unremovable phases in Yukawa matrices for an arbitrary number of fermion 
generations, $N_G$.  A consideration of general $N_G$ yields further insight 
into the physical case $N_G=3$.  However, the
results in this section will not be used directly in our analysis of realistic
models, so that it may be skipped by the reader who is mainly interested in
phenomenological applications. 

     It is interesting to consider the case of Yukawa matrices with all
arbitrary nonzero elements, as we did before, in the context of our first
theorem on the number of unremovable phases.  Just as this case clearly yields
the maximal number of unremovable phases, so also it yields the maximal 
number of complex invariants.  We have proved above that in Corollary 2 that in
this case the 
complex quartic invariants form a complete set, in the sense that the
arguments of the order $2n$ invariants for $2n \ge 6$ can be expressed in terms
of those of the quartic invariants.  

   We now calculate the maximal number of (independent) complex quartic 
invariants.  This is done, as explained above, by performing the calculation in
the case where the $Y^{(f)}$ have all nonzero arbitrary (complex) elements,
since this case clearly yields the maximal number of such invariants. 
Our first result is that the maximal number of complex quartic $P^{(f)}$ 
invariants of each type $f$ ($=u$ or $d$) is
\beq
(N_{P_4,f})_{max} = (^{N_G}_2)^2
\label{np4max}
\eeq
{\it Proof:} \newline
For a given $f=u$ or $d$, the quartic $P^{(f)}$ invariants are
obtained by choosing all rectangles (some of which will be square).  To pick a
rectangle, we pick any two different columns of $Y^{(f)}$; this can be done 
in $(^{N_G}_2)$ ways.  (Note that if we were to choose the two columns to
coincide, we would get a doubly occupied line segment, corresponding to a real
invariant.)  These columns are equivalent (whence the division by $2!$) because
of the fact, noted above, that since each complex invariant and its complex
conjugate yield the same phase constraint, we can count rectangles of the form
${^{\odot}_{\otimes}}{^{\otimes}_{\odot}} = \uparrow \downarrow$ and 
${^{\otimes}_{\odot}}{^{\odot}_{\otimes}}=\downarrow \uparrow$ as equivalent. 
Within one of the columns we pick any two different elements; this can be 
done in any of $(^{N_G}_2)$ ways.  Therefore, the number of (independent) 
complex quartic $P^{(f)}$ invariants of each type $f$ for this case, which is
the maximal number of such invariants, is as given in (\ref{np4max}). $\Box$

    Since the maximal number of quartic $P^{(f)}$ complex invariants of each 
type $f=u$ or $d$, $N_{P_4,f,max}$, is the same for $f=u$ and $f=d$, the 
total maximal number of quartic $P$ complex invariants is just double the
right-hand side of (\ref{npmax}), i.e. 
\beq
(N_{P_4})_{max} = \frac{[N_G(N_G-1)]^2}{2}
\label{np4udmax}
\eeq

    For the maximal number of complex quartic $Q$ invariants, we find
\beq
(N_{Q_4})_{max} = \frac{N_G^3(N_G-1)}{2}
\label{nq4max}
\eeq
{\it Proof:} \newline
As above, the maximal value is calculated by considering the case where
$Y^{(u)}$ and $Y^{(d)}$ have all nonzero arbitrary complex elements. 
First, we pick any column in, say, $Y^{(u)}$; this can be done in $N_G$ ways.
Within this column we choose two different elements, which can be done in
$(^{N_G}_2)$ ways.  This determines one side of the rectangle.  To pick the
other side, we choose any column of $Y^{(d)}$; this can be done in $N_G$ ways.
The two elements within this column are, of course, already determined by those
in the column of $Y^{(u)}$.  As before, the cases $\uparrow \downarrow$ and 
$\downarrow \uparrow$ give complex conjugate invariants and hence the same
phase constraint, so that one does not have to treat them together, which is
equivalent to ignoring an overall reversal in the directions of all arrows.
Hence, the number of complex quartic $Q$ invariants is $N_G(^{N_G}_2)N_G$, i.e.
the result given in (\ref{np4max}). $\Box$

Hence, finally, the maximal value of the total number of independent complex 
quartic invariants is
\beqs
(N_{inv,4})_{max} & = & (N_{P_4})_{max} + (N_{Q_4})_{max} \nonumber \\
                                               \nonumber \\
               & = & \frac{N_G^2(N_G-1)(2N_G-1)}{2}
\label{ninv4max}
\eeqs
For the specific cases $N_G=$1, 2, and 3, we thus have 
\beq
((N_{P_4})_{max}, \ (N_{Q_4})_{max}, \ (N_{inv,4})_{max})=(0,0,0), \ 
(2,4,6), \ (18,27,45) 
\label{numbers}
\eeq

Continuing with the analysis of this case in which all of the elements of
$Y^{(f)}$, $f=u,d$ are nonzero arbitrary complex quantities, we immediately see
that for this case (except for $N_G=1$ for which 
$(N_{inv})_{max}=(N_p)_{max}=0$) the
resultant maximal number of unremovable phases, or, by our theorem on 
invariants, equivalently, the maximal number of independent arguments among 
the complex invariants, $(N_{ia})_{max}=(N_p)_{max}$, is always less than 
$(N_{inv,4})_{max}$, so that the arguments of the various complex quartic 
invariants are not all independent. This is proved easily by recalling our 
earlier result (\ref{npmax}), which, with (\ref{np}), yields 

\beq
(N_{ia})_{max}=(N_p)_{max}=(N_G-1)(2N_G-1)
\label{niamax}
\eeq
Comparison with our result for $(N_{inv,4})_{max}$ in eq. (\ref{ninv4max}) 
shows that 
\beq
(N_{inv,4})_{max} > (N_{ia})_{max} \ \quad for \ N_G \ge 2
\label{comparison}
\eeq
We further recall from Corollary 2 to our theorem on invariants 
that in this case where all elements of the $Y^{(f)}$ are
nonzero arbitrary complex numbers, the arguments of the 6'th and higher order 
complex invariants can be expressed in terms of those of the quartic
invariants.  Although in this case the 6'th order invariants do not yield any
further phase constraints, it is of interest to calculate the maximal number of
these invariants since it shows how this number depends on $N_G$.

    For the number of independent 6'th order complex $P$ invariants in this 
case of $Y^{(f)}$ with all nonzero arbitrary (complex) entries, which 
gives the maximal number of complex $P_6$ invariants, we get
\beq
(N_{P_6,f})_{max}=\frac{[N_G(N_G-1)(N_G-2)]^2}{3!}
\label{np6fmax}
\eeq
{\it Proof:} \newline
We calculate this as follows.  First, we pick a column, in $Y^{(f)}$ ($f=u$ or
$d$), which can be done in $N_G$ ways.  Within this column, we pick two
different elements, which can be done in $(^{N_G}_2)$ ways.  Next, we pick
another column in this matrix, which can be done in $(N_G-1)$ ways. In this
column, we pick the element which is at the same level (i.e. has the same value
of the first index) as the head of the arrow in the first column, or the
element which is at the same level as the tail of this arrow; there is a factor
of 2 corresponding to these two choices.  We then pick a second element in this
column, which is different from the first and is also not at the same level of
the other end of the arrow in the first column (since if it were, we could
immediately form a complex quartic invariant, i.e. we would not be constructing
a genuine 6'th order complex invariant).  There are $(N_G-2)$ ways of choosing
this second element.  Finally, we choose the third column, which is different
from the first two; this can be done in $(N_G-2)$ ways.  The vertical levels of
the elements in this column are now automatically determined from the previous
choices. Since all of the columns are equivalent, we divide by $3!$.  Putting
these factors together gives, for the number of complex 6'th order $P$
invariants, the result $N_G (^{N_G}_2) (N_G-1) \cdot 2 \cdot (N_G-2)
(N_G-2)/3!$, which yields (\ref{np6fmax}). $\Box$ \newline
Including the $f=u$ and $f=d$ sectors, one obtains for the maximal value of
the total number of $P_6$ complex invariants just doubles the right-hand side 
of (\ref{np6fmax}).

For the maximal number of independent 6'th order complex $Q^{(fff')}$ 
invariants, we calculate
\beq
(N_{Q_6,fff'})_{max}=\frac{N_G^3(N_G-1)^2(N_G-2)}{2}
\label{nq6fff'}
\eeq
{\it Proof:} \newline
We consider the $Y^{(f)}$ matrix first.  As before, we pick a column, and two
different elements within this column, which can be done in $N_G(^{N_G}_2)$
ways. We then pick a different column within this matrix; this can be done in
$(N_G-1)$ ways.  In this column 
we pick the element which is at the same level (i.e. has the same value
of the first index) as the head of the arrow in the first column, or the
element which is at the same level as the tail of this arrow; there is a factor
of 2 corresponding to these two choices.  We then pick a second element in this
column, which is different from the first and is also not at the same level of
the other end of the arrow in the first column (since if it were, we could
immediately form a complex quartic invariant, i.e. we would not be constructing
a genuine 6'th order complex invariant).  There are $(N_G-2)$ ways of choosing
this second element.  These two columns are equivalent, so we divide by a
factor of $2!$.  Finally, we pick a column in the other matrix, $Y^{(f')}$,
which can be done in $N_G$ ways.  The elements in this column are now
completely determined by the previous choices.   This yields the result 
$N_G(^{N_G}_2)(N_G-1)\cdot 2 \cdot (N_G-2) \cdot (2!)^{-1} \cdot N_G$, i.e. the
number in (\ref{nq6fff'}). $\Box$

    Including both the $fff'=uud$ and $ddu$ terms just doubles the number in
(\ref{nq6fff'}).  Hence, for the total number of independent complex 6'th order
invariants in this case, we finally get 
\beqs
(N_{inv,6})_{max} & = & 2[(N_{P_6,f})_{max}+(N_{Q_6,fff'})_{max}] \nonumber \\
                                                      \nonumber \\
              & = & \frac{2N_G^2(N_G-1)^2(N_G-2)(2N_G-1)}{3}
\label{ninv6max}
\eeqs
Of course, in the case where the $Y^{(f)}$ have all nonzero arbitrary
complex elements, we have already shown that the quartic complex invariants
form a complete set, so that these 6'th order invariants do not yield any new
phase constraints.  One use of the formulas 
(\ref{np6fmax})-(\ref{ninv6max}) is that they show that there are no complex
6'th order invariants for $N_G=2$.  We will show below that all higher order 
complex invariants also vanish for $N_G=2$. 

    The maximal number of independent complex invariants of order $\ge 8$ may
be calculated in a similar manner.  Here, as explained before, we consider only
irreducible 8'th order invariants, i.e. those 
 which cannot be factored into products of 
lower-order invariants. Thus, for example, 8'th order invariants formed as 
the products of two different quartic invariants are not counted.  Since 
the proofs become rather long, we will simply give the results. 
For the maximal number of (irreducible) 8'th order $P^{(f)}$ invariants of 
each type $f=u$ or $d$ we find
\beq
(N_{P_8,f})_{max}=\frac{N_G^2(N_G-1)^2(N_G-2)(N_G-3)^3}{3!}
\label{np8fmax}
\eeq
For the maximal number of complex $Q^{(3,1)}_8$ invariants (which is, of
course, equal to that of the $Q^{(1,3)}_8$ invariants), we calculate 
\beq
(N_{Q^{(3,1)}_8})_{max}=(N_{Q^{(1,3)}_8})_{max}=
\frac{N_G^3(N_G-1)^2(N_G-2)(N_G-3)^2}{3}
\label{nq31max}
\eeq
Finally, for the maximal number of complex $Q^{(2,2)}_8$ invariants, we get
\beq
(N_{Q^{(2,2)}_8})_{max}=\frac{N_G^3(N_G-1)^3(N_G-2)(N_G-3)}{4}
\label{nq22max}
\eeq
We now observe that the maximal numbers of each of these various types of 
8'th order invariants, $(N_{P_8,f})_{max}$ in (\ref{np8fmax}),
$(N_{Q^{(3,1)}_8})_{max}=(N_{Q^{(1,3)}_8})_{max}$ in (\ref{nq31max}), and 
$(N_{Q^{(2,2)}_8})_{max}$ in (\ref{nq22max}), all vanish unless 
$N_G \ge 4$.  This
explicit calculation thus provides an illustration of the general result in
part (d) of our theorem on invariants.  Recall that that result applied more
generally to all complex invariants, including reducible ones, and stated that
for the physical case $N_G=3$ the arguments of any complex invariant of order
$\ge 8$ could be expressed in terms of the arguments of complex invariants of
order 4 and 6.  Our general calculation above shows that, for the physical 
case $N_G=3$, there are no irreducible complex invariants of order 8, i.e. all
8'th order complex invariants are products of lower-order invariants, and
hence obviously their arguments can be expressed in terms of those of these
lower-order complex invariants.

    One might be curious how part (d) would read if $N_G$ were equal to 2.  
The answer is:
\newline
\noindent {\it Theorem:} \newline
For $N_G=2$, the only (irreducible) complex invariants are of 4'th order. 
\newline
{\it Proof:} \newline
To prove this, we calculate the maximal number of complex invariants as before,
by using the case where the $Y^{(f)}$ have all nonzero arbitrary complex
elements. 
Consider first a $P^{(f)}_{2n}$ complex invariant of order $2n \ge 6$.  To 
form this, one must pick $n \ge 3$ different columns from $Y^{(f)}$.  But 
this is impossible, since $Y^{(f)}$ has only $N_G=2$ columns.  Consider next 
a 6'th order complex $Q$-type invariant, say $Q^{(2,1)_6}$.  To form this, one
must pick two columns from $Y^{(u)}$ with the property that the ends of the
arrows do not match each other (if they did, one would form a $P_4$ invariant).
But this is impossible since $Y^{(u)}$ is just a $2 \times 2$ matrix.  Similar
reasoning shows that it is also impossible to form any higher-order irreducible
complex invariant. $\Box$   \newline
Of course, there may be reducible complex
invariants of higher order, e.g., products of two different quartic invariants.

For this case $N_G=2$ it is easy to construct all such invariants.  To obtain
the maximal set, we assume that the $Y^{(f)}$ have all nonzero arbitrary
(complex) elements; obviously, if some elements are zero, the resultant set of
complex invariants would be commensurately reduced.   According
to our general result (\ref{np4max}), $(N_{P_4,f})_{max}=1$, and the explicit 
complex invariants are
\beq
P^{(f)}_{11,22}=Y^{(f)}_{11}Y^{(f)}_{22}Y^{(f)*}_{21}Y^{(f)*}_{12} 
\label{pf1122}
\eeq
for $f=u,d$. From our general result (\ref{nq4max}) it follows that there are 
$(N_{Q_4})_{max}=4$ independent complex quartic $Q$ invariants.  These are 
\beq
Q_{11,22}=Y^{(u)}_{11}Y^{(d)}_{22}Y^{(u)*}_{21}Y^{(d)*}_{12} 
\label{q1122}
\eeq
\beq
Q_{12,21}=Y^{(u)}_{12}Y^{(d)}_{21}Y^{(u)*}_{22}Y^{(d)*}_{11} 
\label{q1221}
\eeq
\beq
Q_{11,21}=Y^{(u)}_{11}Y^{(d)}_{21}Y^{(u)*}_{21}Y^{(d)*}_{11} 
\label{q1121}
\eeq
and
\beq
Q_{12,22}=Y^{(u)}_{12}Y^{(d)}_{22}Y^{(u)*}_{22}Y^{(d)*}_{12} 
\label{q1222}
\eeq
From eq. (\ref{npmax}), we know that there are $N_p=3$ unremovable phases for
this case, and correspondingly, among the $N_{inv}=6$ complex (quartic)
invariants there are $N_{ia}=N_p=3$ invariants with linearly independent
arguments.  These may, for example, be taken to be $P^{(u)}_{11,22}$,
$P^{(d)}_{11,22}$, and $Q_{11,22}$.

    Returning to the case of general $N_G$, as one makes some of the 
elements of the $Y^{(u)}$ and $Y^{(d)}$ zero, the number of unremovable 
phases and the number of complex invariants are decreased.  The 
amount by which the number of unremovable phases is decreased is not solely a
function of the number of zeroes, but in addition depends on their locations. 
For almost all of the models which we have studied, one finds a simple
empirical relation $N_p = (N_{p})_{max}-N_z = (2N_G-1)(N_G-1)-N_z$, where we
recall that $N_z$ is the total number of zero elements in $Y^{(u)}$ and 
$Y^{(d)}$.  This is not a general result; in section 5 we exhibit a toy model
(no. 8) which, considered as an $N_G=3$ model, is an exception.  
However, that model is degenerate, in the 
sense that it involves a decoupled third generation in both up and down quark
sectors. If one considers only the mutually coupled sector of the model, then
it reduces to a $N_G=2$ model, and does obey the $N_G=2$ version of the above 
empirical relation.

    As a final item in this section, we contrast the unremovable and hence 
physically meaningful phases in the quark Yukawa (or mass) matrices, 
with the unremovable phase(s) in the Cabibbo-Kobayashi-Maskawa (CKM) quark 
mixing matrix, $V$, defined by the charged weak current 
\beq
J^{\mu} = \bar u_{j L, m}\gamma^\mu V_{jk} d_{k L, m}
\label{ckm}
\eeq
where $u_{j L,m}$ and $d_{k L,m}$ are the left-handed components of the mass
eigenstates of the $Q=2/3$ and $Q=-1/3$ quark fields (indicated by the 
subscript $m$) defined after the spontaneous symmetry breaking of $SU(2) \times
U(1)$, with $u_{1 L,m}=u_{L,m}$, $u_{2L,m}=c_m$, $u_{3L,m}=t_{L,m}$, and
similarly for $d_{jL,m}$. If the mass matrices are diagonalized by the 
biunitary transformations 
$U_{f,L}Y^{(f)}U_{f,R}^\dagger = Y^{(f)}_{diag}$ for $f=u,d$, then 
$V=U_{u,L}U_{d,L}^\dagger$.  The rephasing properties of $V$ are quite 
different from those of the $Y^{(f)}$, as is obvious from the fact that the
charged weak current is chirality-preserving, connecting left-handed with
left-handed chiral components of individual physical mass eigenstates of 
quark fields, whereas the Yukawa terms connect left-handed with 
right-handed chiral components.  Furthermore, the Yukawa matrices, being 
defined via $SU(3) \times SU(2) \times U(1)$-invariant Yukawa couplings, 
involve the actual $SU(3) \times SU(2) \times U(1)$ fields, i.e. the $N_G$ 
left-handed SU(2) doublets $Q_{jL}$ and the $2N_G$ right-handed SU(2) 
singlets $u_{jR}$ and $d_{jR}$ (not mass eigenstates).  The rephasing equations
for the Yukawa matrices were given above in (\ref{qrephase})-(\ref{drephase}).
In contrast, when one determines the unremovable phases in $V$, since it is
defined in terms of the mass eigenstates themselves, the rephasing equations
which one uses are
\begin{eqnarray}
u_{j,L m} & = & e^{i\theta_j} u_{j,L m}' \nonumber \\
d_{j,L m} & = & e^{i\phi_j} d_{j,L m}'
\label{ud}
\end{eqnarray}
for $j=1,...,N_G$. 
The number of these rephasing equations is $2N_G$, as opposed to the $3N_G$
rephasing equations (\ref{qrephase})-(\ref{drephase}) for the Yukawa matrices. 
The resultant CKM matrix is 
\beq
V'_{jk}=e^{i(-\theta_j + \phi_k)}V_{jk}
\label{vrephased}
\eeq
Recall the well-known argument that since (i) $V$ is unitary, and hence 
specified by $N_G^2$ real parameters, (ii) there are $2N_G$ rephasings as 
noted; and (iii) the overall rephasing U(1) defined by (\ref{ud}) with 
$\theta_j=\phi_j$ for all $j=1,...,N_G$ leaves $V$ invariant, it follows that 
there are $N_G^2-(2N_G-1)=(N_G-1)^2$ real parameters specifying $V$, of which
$N_G(N_G+1)/2$ are rotation angles and the remainder are 
\beq
N_{p,V}=\frac{(N_G-1)(N_G-2)}{2}
\label{npv}
\eeq
unremovable, physically meaningful, phases.  An immediate contrast is that in
general, $N_{p,V}$ is only a function of $N_G$, whereas the number of
unremovable phases in the Yukawa matrices do not just depend on $N_G$ but
rather on the detailed structure of these matrices.  An important inequality is
that the number of unremovable phases in the quark Yukawa matrices is at least
as large as the number of unremovable phases in the CKM matrix $V$:
\beq
N_p \ge N_{p,V}
\label{npnpv}
\eeq
The proof of this is obvious since the phases in $V$ originated, as $V$ itself
did, from the diagonalization of the Yukawa (equivalently, mass) matrices as
defined in (\ref{massterm}). Our discussion of specific realistic models will 
illustrate this inequality; in particular, in the physically interesting case 
$N_G=3$, the CKM matrix has $N_{p,V}=1$ unremovable (CP-violating) phase, 
whereas the realistic 
models typically have at least $N_p=2$ unremovable phases in the Yukawa 
matrices.  From our inequality (\ref{npnpv}), there follows the theorem that 
phases in quark mass matrices do not necessarily violate CP.  The issue of 
CP violation involving Yukawa couplings also depends on how many Higgs are
present in the theory and how they couple to the quarks.  In order to make our
analysis as general as possible, we have defined the mass and Yukawa matrices
as in eq. (\ref{massterm}) without explicitly specifying the precise Higgs 
content of the theory; however, most recent studies of actual models (including
our own) assume the minimal supersymmetric standard model, for which the Higgs
were listed in (\ref{h1vev}) and (\ref{h2vev}).  For a typical case where 
$N_p > N_{p,V}$, i.e. $N_p > 1$ for the $N_G=3$ case of physical interest, 
it follows that some phases in the quark mass matrices may contribute to
CP-conserving quantities.   In the mathematical case
$N_G=1$, eqs. (\ref{npmax}) and (\ref{npv}) imply that $(N_p)_{max}=N_{p,V}=0$.
For the case $N_G=2$, the CKM matrix still has no phase, but (\ref{npmax})
shows that $N_p$ may be as large as 3.  Indeed, in the specific $N_G=2$ case
worked out completely in eqs. (\ref{pf1122})-(\ref{q1222}), $N_p$ is equal to
3. Thus, in this $N_G=2$ case, the phases in the Yukawa (mass) matrices do 
not yield any phase in the CKM matrix.

\section{Applications to Specific Models}
\label{models}

    Although our theorems are quite general, it is useful to see how they apply
to various specific models.  We shall do this in the present section.  Over the
years, a large number of models have been proposed for the origin of quark
masses and mixing.  In some of the earliest efforts, the forms of the models 
were hypothesized to apply near to the electroweak mass scale.  In general,
various studies can be classified according to whether they assume a
theoretical framework of perturbative or nonperturbative electroweak symmetry 
breaking.  We shall concentrate here on the class of models in which
the observed electroweak symmetry breaking is perturbative.  In this class, 
an appealing framework is provided by supersymmetric extensions of the 
standard model, which stabilize the Higgs sector and hence stabilize the 
hierarchy between the electroweak scale and the Planck scale.  Among
these supersymmetric extensions, the minimal supersymmetric standard model
(MSSM) has the particular feature that it yields gauge coupling unification,
thereby providing a self-consistent basis for a supersymmetric grand unified
theory (SUSY GUT) \cite{susy}.  In turn, for studies of fermion masses and 
mixing, a (supersymmetric) grand unified theory has the advantages that 
(i) one can naturally relate quark and lepton masses; (ii) with the grand 
unified group SO(10) one can naturally obtain (complex) symmetric Yukawa 
matrices at the GUT scale, which helps to minimize parameters. 
Thus, a number of studies have been carried out of particular 
(complex) symmetric forms for Yukawa matrices at the (supersymmetric) 
grand unified scale, in which the renormalization group equations 
of the MSSM are used to evolve these forms down to the electroweak 
scale, where they are diagonalized to yield the known quark masses and the 
CKM quark mixing matrix $V$. In these studies, one assumes some symmetries 
(which are usually discrete in current work) to prevent various Yukawa 
couplings and thereby render various entries in the Yukawa matrices zero.  
The purpose of this is, of course, to minimize the number of parameters and 
hence increase the apparent predictiveness.  (Parenthetically, 
we recall that when one evolves these forms for the Yukawa
matrices down to the electroweak level, they do not, in general, remain
precisely symmetric, and some of the zero entries can become nonzero.) These
studies are phenomenological in the sense that they do not explain the origin
of these discrete symmetries.  The most likely origin of the 
discrete symmetries is probably an underlying string theory (in
which these are actually local symmetries).  However, ironically, 
string theories do not generically yield either simple gauge 
groups (the essence of grand unification) or symmetric (complex or real) mass 
matrices, at least at the present limited level of understanding of their
physical properties \cite{strings}.  
At a phenomenological level it is reasonable to
seek forms for the $Y^{(f)}$ with as many zero entries as possible.  However,
if and when one understands fermion masses and mixing in terms of a truly
fundamental theory, with perhaps no free parameters, then since everything is
(at least in principle) calculable, it makes no difference whether the elements
of the $Y^{(f)}$ are exactly zero or just small.  The entries which are
modelled as being exactly zero might well be nonzero, calculable quantities
suppressed by sufficiently high powers of $(M_r/\bar M_P)$, where $M_r$ is 
the reference scale at which one analyzes the effective Yukawa interaction.
Thus, the real goal is not necessarily to maximize the number of zero entries
in the Yukawa matrices of a viable model as one necessarily tries to in a
phenomenological approach; rather, it is to gain an understanding of the
(presumably unique) fundamental theory, together with a calculational 
ability which would enable one to calculate these matrices from first 
principles.  Given the widely acknowledged \cite{strings} limitations on one's
understanding of the phenomenological consequences of strings and, at the level
of this understanding, the apparent lack of uniqueness of string theory, as 
evident, e.g., in the 4D free-fermion constructions, it does not appear that
one is very close to achieving this goal at present. Finally, for
completeness, we must note that there are a number of concerns which 
must be faced in actual model-building with supersymmetric theories, such as,
for example, the ``$\mu$ problem'' and the problem of splitting light Higgs 
doublets from other components of grand unified Higgs representations 
which must be heavy, and so forth; these are reviewed, e.g., in \cite{susy}.

   With these brief theoretical comments, we proceed to illustrate our general 
theorems with specific models.  We emphasize that our purpose here is not to
propose new viable models of quark Yukawa matrices with the attendant
renormalization group calculations and comparison with experiment; rather, our
purpose is to consider (generalizations of) forms which are known to be
experimentally viable and, for these, to determine the unremovable phases and
which elements can be made real by rephasings. 
Since our results do not depend on whether or
not the Yukawa matrices are symmetric at some mass scale, we shall consider the
general case of non-symmetric Yukawa matrices.\footnote{From a SUSY GUT 
model-builder's point of view, this might seem
perverse, since symmetric Yukawa matrices have the
appeal of reducing the number of parameters and, moreover, can be naturally
obtained in grand unified theories.  But this appeal is offset by one's
awareness of the results of string-based studies, mentioned above, which 
suggest that nature may not, in fact, be described by either a grand unified 
group or by symmetric Yukawa matrices.}  As a basis for this, we will use a
valuable recent study of SUSY GUT Yukawa matrices with the restriction that 
$|Y^{(f)}_{jk}|=|Y^{(f)}_{kj}|$ \cite{rrr}.  This work presented an exhaustive
list of five viable models, with this restriction, 
which we label $M_j$, $j=1-5$.  For the reasons
mentioned above, we shall actually consider the generalizations of these 
models in which $|Y^{(f)}_{jk}|$ is not necessarily equal to $|Y^{(f)}_{kj}|$. 
Again, we emphasize that this does not affect the counting of the number of
unremovable phases or the determination of which elements can be made real by
rephasing.  We label our five generalizations as $M_j'$, $j=1-5$ and discuss 
them in the next five subsections. 
Since the respective special cases are experimentally viable, 
it follows, ${\it a \ fortiori}$, that the generalizations are also viable, 
i.e.,  for appropriate choices of the parameters, one finds 
predictions in agreement with experimental constraints on the CKM 
matrix for experimentally acceptable input values of the top quark mass.  
Our set of models is not intended to be exhaustive, since a comprehensive 
study of viable non-symmetric Yukawa matrices has not been performed (for the 
understandable reason that at a phenomenological level such models have too 
many parameters to make useful predictions).  Model 6 is related to a model
with non-symmetric Yukawa matrices recently sketched in a string
context\cite{faraggi}. 
Models 7 and 8 are toy examples to illustrate certain theoretical points. 

\subsection{Model 1}

   The first illustrative model is defined by the Yukawa matrices
\beq
Y^{(u)} =  \left (\begin{array}{ccc}
                  0 & A_{12} & 0 \\
                  A_{21} & A_{22} & 0 \\
                  0 &    0       &  A_{33} \end{array}   \right  )
\label{yu1}
\eeq
\smallskip
\beq
Y^{(d)} =     \left ( \begin{array}{ccc}
                  0 &  B_{12} &  0 \\
                  B_{21} & B_{22} & B_{23} \\
                  0 & B_{32} & B_{33} \end{array}  \right )
\label{yd1}
\eeq
where here, and below, each element in the Yukawa matrices is, in general, 
complex, since no symmetry requires it to be real.  This model has $N_{z,u}=5$,
$N_{z,d}=3$, $N_z=8$, and $N_{eq}=10$. In the physical context, the forms in
eqs. (\ref{yu1}) and (\ref{yd1}) are 
assumed to hold at a given mass scale, such as the SUSY GUT mass scale 
$M_{GUT} = (1-2) \times 10^{16}$ GeV where the gauge couplings coincide in the
simple MSSM calculation, or a higher mass scale near to the (reduced) Planck
mass $\bar M_P =  2.4 \times 10^{18}$ GeV, which would be expected in
supergravity and string theories, in which one would extend the MSSM
calculation to take account of string threshold corrections and additional
fields which could be present.  Special cases of
this model have been studied in Refs. \cite{rrr} and \cite{ksmodel}.  Besides
the assumption made in both of these works, that 
$|Y^{(f)}_{jk}|=|Y^{(f)}_{kj}|$, $f=u,d$, Ref. \cite{ksmodel} also took 
$|B_{22}|=|B_{23}|$.

   We will give a detailed, pedagogical discussion of the application of our 
methods for this model. 
We first calculate the $T$ matrix.  The reader may recall
here the definition in eq. (\ref{teq}) and the fact that the 
ordering of the columns of $T$ is determined by the definition of the 
vector $v$ in (\ref{vvector}) while the ordering of the equations 
in (\ref{yfrephaseq}), and hence the ordering of the elements of the vector $w$
and the rows of $T$, is defined by the sequence of nonzero elements 
$Y^{(f)}_{jk}$ in (\ref{wvector}) with 
$jk=\{11,12,13,21,22,23,31,32,33\}$ for $f=u$ and then $f=d$.  Explicitly, here
the rows correspond to the rephasing equations for $Y^{(u)}_{12}$, 
$Y^{(u)}_{21}$, $Y^{(u)}_{22}$, $Y^{(u)}_{33}$, $Y^{(d)}_{12}$, 
$Y^{(d)}_{21}$, $Y^{(d)}_{22}$, $Y^{(d)}_{23}$, $Y^{(d)}_{32}$, and 
$Y^{(d)}_{33}$.  We obtain the $N_{eq} \times 3N_G=$ 10-row by 9-column matrix
\beq
T_1 =  \left (\begin{array}{ccccccccc}
     1 & 0 & 0 & 0 & 1 & 0 & 0 & 0 & 0 \\
     0 & 1 & 0 & 1 & 0 & 0 & 0 & 0 & 0 \\
     0 & 1 & 0 & 0 & 1 & 0 & 0 & 0 & 0 \\
     0 & 0 & 1 & 0 & 0 & 1 & 0 & 0 & 0 \\
     1 & 0 & 0 & 0 & 0 & 0 & 0 & 1 & 0 \\
     0 & 1 & 0 & 0 & 0 & 0 & 1 & 0 & 0 \\
     0 & 1 & 0 & 0 & 0 & 0 & 0 & 1 & 0 \\
     0 & 1 & 0 & 0 & 0 & 0 & 0 & 0 & 1 \\
     0 & 0 & 1 & 0 & 0 & 0 & 0 & 1 & 0 \\
     0 & 0 & 1 & 0 & 0 & 0 & 0 & 0 & 1 \end{array} \right )
\label{t1}
\eeq
where the subscript refers to the model.  Further, we calculate the rank of
this matrix, $rank(T_1)=8$.  We can then apply our theorem in (\ref{np}) to
conclude that this model has $N_p = N_{eq}-rank(T)=2$ unremovable, and hence
physically meaningful, phases in the Yukawa matrices $Y^{(f)}$ (and thus of
course also in the mass matrices).  According to our theorem on invariants,
there are therefore $N_{ia}=N_p=2$ complex invariants with independent 
arguments.  For this model we find that $N_{inv,4}=N_{ia}$, i.e. the number 
of quartic complex invariants is equal to the number of complex invariants 
with independent arguments, so that the two complex quartic invariants
fully account for the two unremovable phases and corresponding phase 
constraints.  Explicitly, we find the (nonzero, independent) complex quartic 
invariants 
\beqs
P^{(d)}_{22,33} & = & Y^{(d)}_{22}Y^{(d)}_{33}Y^{(f)*}_{32}Y^{(f)*}_{23} 
\nonumber \\
& = & B_{22}B_{33}B_{32}^*B_{23}^*
\label{pd2233}
\eeqs
and 
\beqs
Q_{12,22} & = & Y^{(u)}_{12}Y^{(d)}_{22}Y^{(u)*}_{22}Y^{(d)*}_{12} 
\nonumber \\
& = & A_{12}B_{22}A_{22}^*B_{12}^*
\label{q1222mod1}
\eeqs
The graphical representations of these complex invariants were given above in 
Fig. 1(a) and Fig. 1(b), respectively. Associated with these are their two 
rephasing-invariant arguments (phases), $\arg(P^{(d)}_{22,33})$ and
$\arg(Q_{12,22})$.\footnote{In the model of Ref. \cite{ks} the phase 
$\theta_{Q_{12,22}}=\arg({Q_{12,22}})$ did not significantly help the fit to 
experimental data and hence was taken to be zero to minimize the number of 
parameters.}
These complex invariants and associated invariant arguments imply constraints 
on which elements of $Y^{(u)}$ and $Y^{(d)}$ can be made real.  Since, 
in general, $\arg(P^{(d)}_{22,33}) \ne 0,\pi$, it follows that (i)
at least one of the $N_p=2$ unremovable phases must 
reside among the set
\beq
S_{P^{(d)}_{22,33}} = \{B_{22},B_{23},B_{32},B_{33}\}
\label{setp2233}
\eeq
and (ii) the $2 \times 2$ submatrix in $Y^{(d)}$ formed by the set
(\ref{setp2233}), 
\beq
           \left ( \begin{array}{cc}
                  B_{22} & B_{23} \\
                  B_{32} & B_{33} \end{array}  \right )
\label{2x2sub}
\eeq
and hence also $Y^{(d)}$ itself, cannot be made real or hermitian.\footnote{
As noted in Ref. \cite{ks}, our phase results disagree with those in Ref. 
\cite{rrr} for the five models which they consider.  In particular, we 
disagree with (i) their statement (footnote on p. 21
of Ref. \cite{rrr}) that (complex) $3 \times 3$ symmetric mass or Yukawa 
matrices can be transformed to hermitian form by rephasings of the quark 
fields; (ii) the counting of unremovable
phases in the Yukawa matrices in that work; and (iii) the claims made
concerning which elements of these matrices can be made real by rephasings of
fermion fields.  For the models $M_j'$, $j=1-5$ considered here, and hence also
for the corresponding models $M_j$ of Ref. \cite{rrr}, we find the values of 
$N_p$ are respectively $2,3,2,2,2$, not $1,2,1,1,1$ as claimed there.  In Ref.
\cite{rrr} the rephased forms of the Yukawa matrices (their Table 1) list
$Y^{(d)}$ as being hermitian and $Y^{(u)}$ as being real for all five models.
We find that $Y^{(d)}$ cannot in general be made hermitian (nor can the 
submatrix (\ref{2x2sub}) be made real or hermitian) in models 1, 2, and 3.  We
also find that $Y^{(u)}$ cannot be made real (or hermitian) in models 4 and 5.
We have communicated our results to P. Ramond and G. G. Ross and thank them 
for discussions.  Our differences do not invalidate the important 
conclusions of Ref. \cite{rrr} that these five forms are experimentally 
viable, since the inclusion of more phases can only improve the fit to 
experiment.} 
Second, since, in general, $\arg(Q_{12,22}) \ne 0,\pi$, it follows that the
elements in the set 
\beq
S_{Q_{12,22}} = \{A_{12},A_{22},B_{12},B_{22}\}
\label{setq1222}
\eeq
(iii) cannot in general be made simultaneously real and thus, 
(iv) if one chooses $Y^{(u)}$ real, then it is not possible to make 
$B_{12}$ and $B_{22}$ both real.  These constitute the complete set of 
rephasing constraints on the $Y^{(f)}$, $f=u,d$. These constraints allow both 
phases to be put in $Y^{(d)}$ and both to be put in the set 
$S_{P^{(d)}_{22,33}}$. 
If one chooses to make $B_{22}$ and $B_{12}$ real, then one
cannot make $Y^{(u)}$ real, and must assign one phase to $A_{12}$ or $A_{22}$
and the second to $B_{23}$, $B_{32}$ or $B_{33}$.  The elements $A_{21}$, 
$A_{33}$, and $B_{21}$ do not occur in any complex invariants; consequently,
their phases are unconstrained and may be rephased
to arbitrary values by the rephasing transformations
(\ref{qrephase})-(\ref{drephase}).  
Our results show that it is not, in general, true that a (complex) 
symmetric Yukawa or mass matrix can be transformed to a hermitian matrix by 
the rephasings of fermion fields. 

As an explicit illustration, after rephasings, one could obtain 
\beq
Y^{(u)'} =  \left (\begin{array}{ccc}
                  0 & |A_{12}| & 0 \\
                  |A_{21}| & |A_{22}| & 0 \\
                  0 &    0       &  |A_{33}| \end{array}   \right  )
\label{yu1r}
\eeq
and
\beq
Y^{(d)'} =     \left ( \begin{array}{ccc}
                  0 &  |B_{12}|e^{-i\: \arg(Q_{12,22})} &  0 \\
     |B_{21}| & |B_{22}|e^{i\: \arg(P^{(d)}_{22,33})} & |B_{23}| \\
                  0 & |B_{32}| & |B_{33}| \end{array}  \right )
\label{yd1r}
\eeq
where for any real quantity, $|Y^{(f)}_{jk}|$, one could have obtained 
$-|Y^{(f)}_{jk}|$ due to the freedom of using $\eta^{(f)}_{jk}=1$ rather than 
$\eta^{(f)}_{jk}=0$ in (\ref{yfrephaseq}).  Since
$\arg(Y^{(d)}_{21})=\arg(B_{21})$ is unconstrained, one could also have 
obtained $Y^{(d)'}_{21}=|B_{21}|e^{i \: \arg(Q_{12,22})}$ in eq. (\ref{yd1r}), 
as discussed in
connection with eq. (\ref{yfhermitian}), although of course this choice has no
effect on the physics.  An example of a form which could not, in general, be
obtained from any rephasings of fermion fields is the hermitian form 
\beq
Y^{(d)'} \ne     \left ( \begin{array}{ccc}
                  0 &  |B_{12}|e^{i\theta} &  0 \\
        |B_{21}|e^{-i\theta} & |B_{22}| & |B_{23}|e^{i\phi} \\
                  0 & |B_{32}|e^{-i\phi} & |B_{33}| \end{array}  \right )
\label{yd1rr}
\eeq
for any values of $\theta$ and $\phi$.  As is evident from our invariants, this
statement does not depend on whether or not the initial $Y^{(u)}$ and $Y^{(d)}$
are (complex) symmetric. 

   For this model, we shall also illustrate explicitly the operations which 
enter into the proof of our theorem (\ref{np}).  In order to get the 
invertible matrix $\bar T$, we select and delete $N_{eq}-r_{_T}=10-8=2$ 
rows from the $T$ matrix for this model, given in eq. (\ref{t1}).  
This corresponds to deleting two elements of the vector $w$ 
defined in (\ref{wvector}) to get $\bar w$, i.e. not rephasing the
corresponding $Y^{(f)}_{jk}$.  Given that we have constructed the full set of
$N_{ia}$ invariants with independent phases, we know the full set of allowed
choices for elements to be rephased and correspondingly the choices for the
$N_p=2$ elements which will not be rephased.  In this model, the rephasing
equations for the elements comprising the complex invariants 
$P^{(d)}_{22,33}$ and $Q_{12,22}$ correspond to the rows 
(1,3,5,7) and (7,8,9,10) of the matrix $T$, respectively.  
Hence we must delete one row from the first set and one row
from the second set; i.e. we have the following 15 choices of row pairs to 
delete: $(r1,r2)$, with $r1$ chosen from (1,3,5,7) and $r2$ chosen from
(7,8,9,10), excluding, of course, the choice $r1=r2=7$.  For our specific 
illustration, we delete rows 5 and 7 and thus do not rephase $Y^{(d)}_{12}$ or 
$Y^{(d)}_{22}$.  Note that if one were to try to remove two rows in 
a manner which would violate the phase constraints, one's error would
immediately be signalled by a reduction in the rank of the resultant matrix.
For example, if one were to try to remove rows 1 and 2, one would obtain a 
matrix with rank 7.  Recall that this choice is forbidden since it would entail
the implication that one could rephase all other elements of $Y^{(u)}$ and
$Y^{(d)}$, which is false, since it would mean that one could in general 
rephase the argument of $Q_{12,22}$ to zero, contrary to fact.  Note that,
in principle, one may determine which rows may be deleted simply by testing the
rank of the resultant matrix, without first constructing the complete set of
complex invariants.  However, obviously the more efficient procedure is to
construct this set of invariants first, since once this is done, it is clear
which sets of $N_p$ rows can or cannot be deleted consistent with the phase
constraints.  Continuing with our explicit calculational example, having
deleted the rows 5 and 7 from $T$, and thereby obtained an $r_{_T} \times 
3N_G=$ 8-row by 9-column matrix, we next delete $3N_{G}-r_{_T}=1$ 
column from this matrix in order finally to get the square, 
$r_{_T} \times r_{_T} = 8 \times 8$ matrix $\bar T$. This corresponds to
deleting one element of the vector $v$ defined in eq. (\ref{vvector}) to get 
$\bar v$.  We find that any of the nine columns of $T$ may be deleted in this
step; for definiteness, we pick the ninth.  This yields the reduced matrix
$\bar T$ for this model:
\beq
\bar T_1 =  \left (\begin{array}{cccccccc}
            1 & 0 & 0 & 0 & 1 & 0 & 0 & 0  \\
            0 & 1 & 0 & 1 & 0 & 0 & 0 & 0  \\
            0 & 1 & 0 & 0 & 1 & 0 & 0 & 0  \\
            1 & 0 & 0 & 0 & 0 & 0 & 0 & 1  \\
            0 & 1 & 0 & 0 & 0 & 0 & 0 & 1  \\
            0 & 1 & 0 & 0 & 0 & 0 & 0 & 0  \\
            0 & 0 & 1 & 0 & 0 & 0 & 0 & 1  \\
            0 & 0 & 1 & 0 & 0 & 0 & 0 & 0  \end{array} \right )
\label{t1bar}
\eeq
(with determinant $-1$).  By construction, this is invertible, and we find 
\beq
\bar T_1^{-1} = \left (\begin{array}{cccccccc}
            1 & 0 & -1 & 0 & 0 & 1 & 0 & 0  \\
            0 & 0 & 0 & 0 & 0 & 1 & 0 & 0  \\
            0 & 0 & 0 & 0 & 0 & 0 & 0 & 1  \\
            0 & 1 & 0 & 0 & 0 & -1 & 0 & 0  \\
            0 & 0 & 1 & 0 & 0 & -1 & 0 & 0  \\
            0 & 0 & 0 & 1 & 0 & 0 & 0 & -1  \\
            0 & 0 & 0 & 0 & 1 & -1 & 0 & 0  \\
            0 & 0 & 0 & 0 & 0 & 0 & 1 & -1  \end{array} \right )
\label{t1barinv}
\eeq

    As an illustration of our theorem on invariants, we discuss the 6'th
order complex invariants for this model.  Since we have already constructed as
many independent complex invariants with independent arguments as there are
unremovable phases, using quartic complex invariants, our theorem on invariants
and its specific Corollary 1 imply 
that the argument of any 6'th, as well as higher order, complex invariant, 
are expressible in terms of the arguments of the quartic invariants.  By 
explicit computation, we find that there is one complex 6'th order invariant
(i.e. $N_{inv,6}=1$): 
\beq
Q^{(ddu)}_{32,23,12}=B_{32}B_{23}A_{12}B_{33}^*B_{12}^*A_{22}^*
\label{qddu322312}
\eeq
Thus the total number of independent complex invariants for this model is
$N_{inv}=N_{inv,4}+N_{inv,6}=3$.  The graphical representation of this 
complex invariant is shown in Fig. 3(a) (where here and elsewhere, lines 
which cross each other at a point where there is no $\odot$ or $\otimes$, 
as in the $Y^{(d)}_{22}$ position, do not mean that the 
invariant contains this element).  Note that this 6'th order invariant 
satisfies the condition (i) of part (d) of the above theorem.  It is thus 
easy to find the explicit reduction; we obtain the factorization
\beqs
\arg(B_{32}B_{23}A_{12}B_{33}^*B_{12}^*A_{22}^*) & = & 
\arg(B_{32}B_{23}A_{12}(B_{22}B_{22}^*)B_{33}^*B_{12}^*A_{22}^*) 
\nonumber \\ 
\nonumber \\
& = & \arg(A_{12}B_{22}A_{22}^*B_{12}^*) + \arg(B_{32}B_{23}B_{22}^*B_{33}^*)
\label{redcal}
\eeqs
whence
\beq
\arg(Q^{(ddu)}_{32,23,12}) = \arg(Q_{12,22})-\arg(P^{(d)}_{22,33})
\label{qddu322312red}
\eeq
Indeed, in this case, we find the factorization
\beq
Q^{(ddu)}_{32,23,12}= |B_{22}|^{-2} Q_{12,22} P^{(d)}_{23,32}
\label{qddu322312fac}
\eeq
Recall here that $P^{(d)}_{23,32}=P^{(d)*}_{22,33}$.  Note also that 
according to our definition given above, $Q^{(ddu)}_{32,23,12}$ is
still an irreducible 6'th order complex invariant because it is not expressible
only as a product of lower-order complex invariants; there is also the
$|B_{22}|^{-2}$ factor. 
Graphically, the insertion of the factor $B_{22}B_{22}^*$ in eq. (\ref{redcal})
is equivalent to inserting a coincident arrow head and tail, or $\odot$ and
$\otimes$ in the $Y^{(d)}_{22}$ position.  Having done this, one sees
immediately that the 6'th order invariant factorizes in the manner specified by
eq. (\ref{qddu322312fac}).  
In addition to using our theorem on invariants to show 
that the argument of $Q^{(ddu)}_{32,23,12}$ can be expressed in terms of those
of quartic complex invariants, another way to see this is to use the $Z$
matrix method discussed above.  For this, we first define the vector $\xi$
defined in eq. (\ref{xidef}) for the present model: 
\beq
(\xi)_1=(\arg(P^{(d)}_{22,33}), \ \arg(Q_{12,22}), \ 
\arg(Q^{(ddu)}_{32,23,12}))^T
\label{xi1}
\eeq
where the subscript on $(\xi)_1$ denotes the fact that this is the $\xi$ vector
for model 1.  The ordering of the elements in the vector $w$ was given in
general after its definition, eq. (\ref{wvector}), and explicitly for this
model.  Then $Z$ is an $N_{inv} \times N_{eq}=$ 3-row by 10-column 
matrix. From the definition (\ref{zeq}) we calculate it to be 
\beq
(Z)_1 =  \left (\begin{array}{cccccccccc}
     0 & 0 & 0 & 0 & 0 & 0 & 1 &-1 &-1 & 1 \\
     1 & 0 &-1 & 0 &-1 & 0 & 1 & 0 & 0 & 0 \\
     1 & 0 &-1 & 0 &-1 & 0 & 0 & 1 & 1 & 1  \end{array} \right )
\label{z1}
\eeq
where again the subscript on $(Z)_1$ refers to model 1.  Since the elements
$A_{21}$, $A_{33}$, and $B_{21}$ do not occur in any complex invariants, they
may be removed from the vector $w$ to yield $w_r$, and correspondingly, one may
remove the second, fourth, and sixth columns from $Z_1$ to form the reduced
4-row by 7-column matrix $Z_r$ defined in (\ref{zreq}), 
\beq
(Z_r)_1 =  \left (\begin{array}{ccccccc}
         0 & 0 & 0 & 1 &-1 &-1 & 1 \\
         1 &-1 &-1 & 1 & 0 & 0 & 0 \\
         1 &-1 &-1 & 0 & 1 & 1 & 1  \end{array} \right )
\label{zr1}
\eeq
Since we have only removed columns of zeroes, this does not change the rank of
the matrix. 
We calculate that $rank((Z)_1)=rank((Z_r)_1)=2$, in agreement with our 
theorem that $N_p=N_{ia}$ and the relation $N_{ia}=rank(Z)$ from eq. 
(\ref{rankz}).  We find further that removing the last row from $(Z)_1$ or
$(Z_r)_1$ does not reduce the rank, which shows 
that the argument of $Q^{(ddu)}_{32,23,12}$ can be expressed in terms
of those of $P^{(d)}_{22,33}$ and $Q_{12,22}$. 
 The phase properties and complex invariants of the Yukawa matrices for this 
and the other realistic models to be discussed are summarized in Table 1. 
Having given a detailed discussion of the details of our analysis for this
model, we next proceed with the others, which we will discuss more briefly. 

\begin{table}
\begin{center}
\begin{tabular}{|c|c|c|c|c|c|c|c|c|} \hline \hline  & & & & & & & & \\
model & $N_{eq}$ & $r_{_T}$ & $N_p=N_{ia}$ & $N_{inv,4}$ & $P_4$ & $Q_4$ & 
$N_{inv,6}$ & $P_6$ or $Q_6$ \\
 & & & & & & & & \\
\hline \hline & & & & & & & & \\
1 & 10 & 8 & 2 & 2 & $P^{(d)}_{22,33}$ & $Q_{12,22}$ & 1 & 
$Q^{(ddu)}_{32,23,12}$ \\ 
 & & & & & & & & \\ \hline  & & & & & & & & \\ 
2 & 11 & 8 & 3 & 4 & $P^{(d)}_{22,33}$ & $Q_{12,32}$,\ $Q_{23,32}$,\ 
$Q_{23,33}$ & 2 & $Q^{(uud)}_{32,23,12}$, \ $Q^{(ddu)}_{22,33,12}$ \\
 & & & & & & & & \\ \hline  & & & & & & & & \\ 
3 & 10 & 8 & 2 & 2 & $P^{(d)}_{2233}$ & $Q_{13,32}$ & 1 & 
$Q^{(ddu)}_{22,33,13}$ \\ 
 & & & & & & & & \\ \hline  & & & & & & & & \\ 
4 & 10 & 8 & 2 & 2 & $P^{(u)}_{22,33}$ & $Q_{12,22}$ & 1 & 
$Q^{(uud)}_{32,23,12}$ \\
 & & & & & & & & \\ \hline  & & & & & & & & \\
5 & 10 & 8 & 2 & 2 & $P^{(u)}_{22,33}$ & $Q_{13,22}$ & 1 & 
$Q^{(uud)}_{32,13,22}$ \\
 & & & & & & & & \\ \hline  & & & & & & & & \\
6 & 12 & 8 & 4 & 7 & $P^{(u)}_{22,33}$ & $Q_{12,22}$, \ $Q_{12,32}$, \ 
$Q_{22,32}$, & 4 & $Q^{(uud)}_{22,33,12}$, \ $Q^{(uud)}_{33,12,22}$ \\ 
 & & & & & & & & \\  
  &    &   &   &   &  $P^{(d)}_{22,33}$ & $Q_{22,33}$, \ $Q_{23,33}$ & & 
$Q^{(ddu)}_{22,33,12}$, \ $Q^{(ddu)}_{33,12,22}$ \\
 & & & & & & & & \\ \hline \hline
\end{tabular}
\end{center}
\caption{Summary of unremovable phases and complex invariants for the six
realistic models of Yukawa matrices. }
\end{table}

\subsection{Model 2}

     This model is defined by 
\beq
Y^{(u)} =  \left (\begin{array}{ccc}
                  0 & A_{12} & 0 \\
                  A_{21} & 0 & A_{23} \\
                  0 & A_{32} & A_{33} \end{array}   \right  )
\label{yu2}
\eeq
and $Y^{(d)}$ as in model 1, i.e. 
\smallskip
\beq
Y^{(d)} =     \left ( \begin{array}{ccc}
                  0 &  B_{12} &  0 \\
                  B_{21} & B_{22} & B_{23} \\
                  0 & B_{32} & B_{33} \end{array}  \right )
\label{yd2}
\eeq 
This model has $N_{z,u}=4$, $N_{z,d}=3$, and hence $N_{eq}=11$.  We calculate 
that $T$ is the $11 \times 9$ matrix
\beq
T_2 =  \left (\begin{array}{ccccccccc}
     1 & 0 & 0 & 0 & 1 & 0 & 0 & 0 & 0 \\
     0 & 1 & 0 & 1 & 0 & 0 & 0 & 0 & 0 \\
     0 & 1 & 0 & 0 & 0 & 1 & 0 & 0 & 0 \\
     0 & 0 & 1 & 0 & 1 & 0 & 0 & 0 & 0 \\
     0 & 0 & 1 & 0 & 0 & 1 & 0 & 0 & 0 \\
     1 & 0 & 0 & 0 & 0 & 0 & 0 & 1 & 0 \\
     0 & 1 & 0 & 0 & 0 & 0 & 1 & 0 & 0 \\
     0 & 1 & 0 & 0 & 0 & 0 & 0 & 1 & 0 \\
     0 & 1 & 0 & 0 & 0 & 0 & 0 & 0 & 1 \\
     0 & 0 & 1 & 0 & 0 & 0 & 0 & 1 & 0 \\
     0 & 0 & 1 & 0 & 0 & 0 & 0 & 0 & 1 \end{array} \right )
\label{t2}
\eeq
with rank 8.  Hence, from eq. (\ref{np}), there are $N_p=N_{eq}-rank(T)=3$ 
unremovable phases in the $Y^{(f)}$, $f=u,d$.  By part (c) of our theorem on 
invariants, there are correspondingly $N_{ia}=N_p=3$ independent phases of 
complex invariants. We actually find that there are $N_{inv,4}=4$ independent 
complex quartic invariants in this model.  
(Recall from eq. (\ref{ninvia}) that in general
there may be more independent complex invariants than independent arguments of
invariants.) The four independent complex quartic invariants are 
$P^{(d)}_{22,33}$ as in model 1, together with 
\beq
Q_{12,32}=A_{12}B_{32}A_{32}^*B_{12}^*
\label{q1232}
\eeq
\beq
Q_{23,32}=A_{23}B_{32}A_{33}^*B_{22}^*
\label{q2332}
\eeq
and
\beq
Q_{23,33}=A_{23}B_{33}A_{33}^*B_{23}^*
\label{q2333}
\eeq 
Of these, the arguments of the first, third, and fourth invariants are 
related according to 
\beq
\arg(P^{(d)}_{22,33})+\arg(Q_{23,32})-\arg(Q_{23,33})=0
\label{mod2rel}
\eeq
so that, as required by our theorem, the number of independent arguments of
invariants is $N_{ia}=3$.  One can also determine the number $N_{ia}$ without
using part (c) of our theorem and Corollary 1 on invariants by calculating 
the matrix $Z$ defined in eq. (\ref{zeq}) and computing its rank.  In the
present case, it is sufficient, and simpler, to deal with the matrix $Z_{4}$
defined in eq. (\ref{z2neq}) for quartic complex invariants.  As defined
in eq. (\ref{xi2n}) for $2n=4$, the vector $\xi_4$ of arguments of quartic 
invariants has dimension $N_{inv,4}=4$ and is 
\beq
(\xi_4)_1 = (\arg(P^{(d)}_{22,33}), \ \arg(Q_{12,32}), \ \arg(Q_{23,32}), \ 
\arg(Q_{23,33}))^T
\label{xi4mod2}
\eeq
where the subscript 2 on $(\xi_4)_2$ refers to model 2. 
Using (\ref{z2neq}) and (\ref{wvector}) with the specified ordering of 
elements in $w_4$ (and $\eta^{(f)}_{jk}=0$ as noted before), we then calculate 
the $N_{inv,4} \times N_{eq}$ = 4-row by 11-column $Z_{4}$ matrix as
\beq
(Z_4)_2 =  \left (\begin{array}{ccccccccccc}
         0 & 0 & 0 & 0 & 0 & 0 & 0 & 1 &-1 &-1 & 1 \\
         1 & 0 & 0 &-1 & 0 &-1 & 0 & 0 & 0 & 1 & 0 \\
         0 & 0 & 1 & 0 &-1 & 0 & 0 &-1 & 0 & 1 & 0 \\
         0 & 0 & 1 & 0 &-1 & 0 & 0 & 0 &-1 & 0 & 1 \end{array} \right )
\label{z42}
\eeq
This matrix has rank 3, so that the three independent arguments among the four
quartic complex invariants account for the full set of three unremovable 
phases in this model.  By selecting and deleting rows of $(Z_4)_2$, one easily 
establishes which of these quartic invariants have linearly dependent
arguments; deleting row 1, 3, or 4 does not reduce the rank, while deleting row
2 yields a matrix with rank 2.  This means that any one of the sets 
$(Q_{12,32}, \ Q_{23,32}, \ Q_{23,33})$, 
$(P^{(d)}_{22,33}, \ Q_{12,32}, \ Q_{23,33})$, and 
$(P^{(d)}_{22,33}, \ Q_{12,32}, \ Q_{23,32})$ constitutes a
complete set of $N_{ia}=3$ complex invariants with linearly independent
arguments, while the arguments of the invariants in the set 
$(P^{(d)}_{22,33}, \ Q_{23,32}, \ Q_{23,33})$ are linearly dependent, in
agreement with the explicit calculation (\ref{mod2rel}). 
We will take the three independent unremovable phases in this model to be 
$\arg(P^{(d)}_{22,33})$, $\arg(Q_{12,32})$, and $\arg(Q_{23,32})$. 
Since the elements $A_{21}$ and $B_{21}$ do not occur in any complex 
invariants, they may be rephased freely, and 
one may also form the vector $w_r$ and corresponding matrix
$Z_{r,4}$ which is just $Z_{4}$ with the second and seventh columns of zeroes
removed.  
The constraints on rephasing from $P^{(d)}_{22,33}$ are (i) and (ii) as in
model 1, and (iii) none of the sets 
\beq
S_{Q_{12,32}} = \{A_{12},A_{32},B_{12},B_{32}\}
\label{setq1232}
\eeq
\beq
S_{Q_{23,32}} = \{A_{23},A_{33},B_{22},B_{32}\}
\label{setq2332}
\eeq
and 
\beq
S_{Q_{23,33}} = \{A_{23},A_{33},B_{23},B_{33}\}
\label{setq2333}
\eeq
can be made simultaneously real.  In particular, if $Y^{(u)}$ is made real,
then none of the sets $\{B_{12},B_{32}\}$, $\{B_{32},B_{22}\}$, and 
$\{B_{23},B_{33}\}$ can be made simultaneously real.  As in model 1, because of
the constraint from $P^{(d)}_{22,33}$, $Y^{(d)}$ cannot in general be made 
real or hermitian.  An example of a rephasing of the Yukawa 
matrices allowed by these constraints is $Y^{(u)}$ real, 
\beq
Y^{(u)'} =  \left (\begin{array}{ccc}
                  0 & |A_{12}| & 0 \\
               |A_{21}| & 0 & |A_{23}| \\
               0 & |A_{32}| & |A_{33}| \end{array}   \right  )
\label{yu2r}
\eeq
and 
\beq
Y^{(d)'} =     \left ( \begin{array}{ccc}
             0 &  |B_{12}|e^{-i \: \arg(Q_{12,32})} &  0 \\
         |B_{21}| & |B_{22}|e^{-i \: \arg(Q_{23,32})} & |B_{23}| \\
 0 & |B_{32}| & |B_{33}|e^{i( \arg(P^{(d)}_{22,33})+\arg(Q_{23,32}))} 
\end{array}  \right )
\label{yd2r}
\eeq
The construction of $\bar T$ from $T$ is performed in a manner similar to that
discussed for model 1.

    As in model 1, since we have already constructed as many independent 
complex invariants with independent arguments as there are unremovable 
phases, $N_{ia}=N_p=3$, using quartic complex invariants, our theorem on 
invariants and its Corollary 1 imply 
that the arguments of any 6'th, as well as higher order, complex invariants, 
are expressible in terms of the arguments of the quartic invariants and hence
yield no new phase constraints.  We find that in the present model there are 
two different 6'th order complex invariants (so $N_{inv,6}=2$):
\beq
Q^{(uud)}_{32,23,12}=A_{32}A_{23}B_{12}A_{33}^*A_{12}^*B_{22}^*
\label{quud322312}
\eeq
and 
\beq
Q^{(ddu)}_{22,33,12}=B_{22}B_{33}A_{12}B_{23}^*B_{12}^*A_{32}^*
\label{qddu223312}
\eeq
The total number of independent complex invariants in this model is therefore 
$N_{inv}=N_{inv,4}+N_{inv,6}=6$.  
The explicit expressions for the arguments of these 6'th order complex
invariants, in terms of quartic complex invariants, are 
\beq
\arg(Q^{(uud)}_{32,23,12})= \arg(Q_{23,32})-\arg(Q_{12,32})
\label{quud322312red}
\eeq
and
\beq
\arg(Q^{(ddu)}_{22,33,12})=\arg(P^{(d)}_{22,33})+\arg(Q_{12,32})
\label{qddu223312red}
\eeq

Again, a different way of showing that the 6'th order complex invariants 
yield no new constraints is by analyzing the full $Z$ matrix for the model.  
The method should be clear from our previous example, so we omit the details.

    Parenthetically, we consider the special case of this model when
$Y^{(d)}_{22}=B_{22}=0$.  Here $Y^{(u)}$ and $Y^{(d)}$ have the same form.
Although current interest is in models of Yukawa matrices which apply at the
SUSY GUT level, it should be noted that the further specialization in which one
takes $|Y^{(f)}_{jk}|=|Y^{(f)}_{kj}|$ is formally the same as one of the 
early works on quark mass matrices \cite{fritzsch}.  Of course in that work 
the form was hypothesized to apply near the electroweak level in a 
(non-supersymmetric) $SU(2)_L \times SU(2)_R \times U(1)$ electroweak gauge 
model, not a SUSY GUT. As is well known, whether one considers the symmetric
form with $B_{22}=0$ to apply at the electroweak 
level or at the SUSY GUT level with MSSM evolution equations, it is 
currently disfavored by experiment, because of the large value which it 
predicts for $V_{cb}$, given the lower bound on top quark mass indicated by 
both direct limits from CDF and D0 \cite{tmass} and by precision fits to 
LEP data \cite{lep}. It is, however, of
historical interest.  In this case, $N_{eq}=10$, $rank(T)=8$, whence $N_p=2$.
Of the four quartic invariants for model 2, two vanish for this special case: 
$P^{(d)}_{22,33}=0$ and $Q_{23,32}=0$.  The other quartic complex invariants,
$Q_{13,32}$ and $Q_{23,33}$, remain, so that $N_{inv,4}=2$.  The phases of
these are independent, so these two quartic invariants account for
both of the two unremovable phases and resultant phase constraints.  The single
6'th order complex invariant for model 2 vanishes for this special case, which
therefore has no nonzero complex 6'th order invariants.

\subsection{Model 3}

    The third model is given by 
\beq
Y^{(u)} =  \left (\begin{array}{ccc}
                  0 & 0 & A_{13} \\
                  0 & A_{22} & 0 \\
                  A_{31} & 0 & A_{33} \end{array}   \right  ) 
\label{yu3}
\eeq
with $Y^{(d)}$ as in model 1:
\beq
Y^{(d)} =     \left ( \begin{array}{ccc}
                  0 &  B_{12} &  0 \\
                  B_{21} & B_{22} & B_{23} \\
                  0 & B_{32} & B_{33} \end{array}  \right )
\label{yd3}
\eeq
Special cases of this model for which $|Y^{(f)}_{jk}|=|Y^{(f)}_{kj}|$ were
studied in Refs. \cite{giudice} and \cite{rrr}; in Ref. \cite{giudice}, the
further assumptions $|A_{22}|=|A_{13}|$ and $|B_{23}|=2|B_{22}|$ were made. 
The calculation of $T$ should be clear from our previous examples, so we just
mention that it is a $10 \times 9$ matrix and we find its rank to be 8.  Hence,
there are $N_p=2$ unremovable phases in the
$Y^{(f)}$, $f=u,d$.  We find $N_{inv,4}$ independent complex quartic 
invariants; these are $P^{(d)}_{22,33}$ as in the previous models, and 
\beq
Q_{13,32}=A_{13}B_{32}A_{33}^*B_{12}^*
\label{q1332}
\eeq
These two complex quartic invariants clearly have independent phases, so that
they fully account for the two unremovable phases in the model: 
$N_{inv,4}=N_{ia}=2$.  The phase constraint from $P^{(d)}_{22,33}$ is the same
as in models 1 and 2 \footnote{Ref. \cite{giudice} actually presents a form in
which the submatrix (\ref{2x2sub}) of $Y^{(d)}$ is real.  Such a form is not, 
in general, possible to achieve by any fermion rephasings, as we have
already discussed above in the context of model 1.}  The phase 
constraint from $Q_{13,32}$ is that the set of elements 
\beq
S_{Q_{13,32}} = \{A_{13},A_{31},B_{12},B_{32}\}
\label{setq1332}
\eeq
cannot be made simultaneously real.  The elements $A_{31}$, $A_{33}$, and 
$B_{21}$ do not occur in any complex invariants and may be rephased freely. 
An example of an allowed form for the
Yukawa matrices after rephasing of fermion fields is $Y^{(u)}$ real, 
\beq
Y^{(u)'} =  \left (\begin{array}{ccc}
                  0 & 0 & |A_{13}| \\
                  0 & |A_{22}| & 0 \\
                  |A_{31}| & 0 & |A_{33}| \end{array}   \right  ) 
\label{yu3r}
\eeq
and 
\beq
Y^{(d)'} =     \left ( \begin{array}{ccc}
          0 &  |B_{12}|e^{-i \: \arg(Q_{13,32})} &  0 \\
     |B_{21}| & |B_{22}|e^{i \: \arg(P^{(d)}_{22,33})} & |B_{23}| \\
               0 & |B_{32}| & |B_{33}| \end{array}  \right )
\label{yd3r}
\eeq
By our theorem on invariants and its specific Corollary 1, any 6'th order
complex invariants have arguments which can be expressed in terms of those 
of the quartic complex invariants.  We find that the model has one 6'th 
order complex invariant (so $N_{inv,6}=1$):
\beq
Q^{(ddu)}_{22,33,13}=B_{22}B_{33}A_{13}B_{23}^*B_{12}^*A_{33}^*
\label{qddu223313}
\eeq
whose argument is expressed in terms of those of the quartic invariants as 
\beq
\arg(Q^{(ddu)}_{22,33,13})=\arg(P^{(d)}_{22,33})+\arg(Q_{13,32})
\label{qddu223313red}
\eeq
The total number of complex invariants is $N_{inv}=N_{inv,4}+N_{inv,6}=3$. 

\subsection{Model 4}

   The fourth model is defined by 
\beq
Y^{(u)} =  \left (\begin{array}{ccc}
                  0 & A_{12} & 0 \\
                  A_{21} & A_{22} & A_{23} \\
                  0 & A_{32} & A_{33} \end{array}   \right  )
\label{yu4}
\eeq
\smallskip
\beq
Y^{(d)} =     \left ( \begin{array}{ccc}
                  0 &  B_{12} &  0 \\
                  B_{21} & B_{22} & 0 \\
                  0 & 0 & B_{33} \end{array}  \right )
\label{yd4}
\eeq
This model can be obtained from model 1 by the interchange of $Y^{(u)}$ and
$Y^{(d)}$.  Using this fact, one can immediately determine all of the phase 
properties and invariants of the model from those of model 1: $N_p=N_{ia}=2$,
and the two independent quartic complex invariants are
\beq
P^{(u)}_{22,33}=A_{22}A_{33}A_{32}^*A_{32}^*
\label{pu2233}
\eeq
and $Q_{12,22}$ as given in (\ref{q1222}).  These fully account for all phase
constraints in the model, which are given by those for model 1 with the
interchange of $Y^{(u)} \leftrightarrow Y^{(d)}$ ($A_{jk} \leftrightarrow
B_{jk}$).  An allowed form for the Yukawa matrices after rephasing is 
\beq
Y^{(u)'} =     \left ( \begin{array}{ccc}
                  0 &  |A_{12}|e^{-i \: \arg(Q_{12,22})} &  0 \\
     |A_{21}| & |A_{22}|e^{i \: \arg(P^{(u)}_{22,33})} & |A_{23}| \\
                  0 & |A_{32}| & |A_{33}| \end{array}  \right )
\label{yu4r}
\eeq
and
\beq
Y^{(d)'} =  \left (\begin{array}{ccc}
                  0 & |B_{12}| & 0 \\
                  |B_{21}| & |B_{22}| & 0 \\
                  0 &    0       &  |B_{33}| \end{array}   \right  )
\label{yd4r}
\eeq
The model has one 6'th order complex invariant, $Q^{(uud)}_{32,23,12}$ (given
explicitly in (\ref{quud322312}) for model 2).  The argument of this 6'th 
order invariant is expressed in terms of those of the two complex quartic 
invariants by the relation 
\beq
\arg(Q^{(uud)}_{32,23,12})= -\arg(Q_{12,22})-\arg(P^{(u)}_{22,33})
\label{q6mod4red}
\eeq
As this illustrates, it is obvious that if the same 6'th order complex
invariant occurs in two different model and if in both cases its argument is
expressible in terms of those of the respective complex quartic invariants for
these models, then these expressions will be different, if, as will generally 
be the case, the complex quartic invariants are different for the two models.
Specifically, one sees that the relation (\ref{q6mod4red}) in the present
model differs from (\ref{qddu322312red}) in model 2. 

   The special case of model 4 in which $|Y^{(f)}_{jk}|=|Y^{(f)}_{kj}|$,
$f=u,d$ and $Y^{(u)}_{22}=A_{22}=0$ \cite{dhr} is not included in ref. 
\cite{rrr} among its experimentally acceptable forms because of the large
values of $|V_{cb}|$ which it yields and, if other quantities are used as
inputs to get a prediction for the top quark mass $m_t$, the high values of
$m_t$ which it predicts. 
However, this special case, and also its generalization with 
$|Y^{(f)}_{jk}|$ not necessarily equal to $|Y^{(f)}_{kj}|$ has several
interesting properties concerning phases and rephasing invariants.  First, 
since $N_{eq}$ is reduced to 9 while $rank(T)$ remains at 8, there is 
only one unremovable phase, $N_p=1$. 
As is evident from eqs. (\ref{pu2233}) and (\ref{q1222}), both of the
quartic complex invariants $P^{(u)}_{22,33}$ and $Q_{12,22}$ vanish, so 
that there are
no quartic complex invariants; $N_{inv,4}=0$.  Therefore, in contrast to the
general model 4 with nonzero $A_{22}$, in this special case, (i) the only 
complex invariant is the 6'th order invariant $Q^{(uud)}_{32,23,12}$ given 
above in (\ref{quud322312}); 
(ii) as is evident from (\ref{quud322312red}), the
argument of this 6'th order invariant cannot be expressed in terms of any lower
complex invariants, since there are none.  Thus this special case provides an
example of a model in which the lowest-order complex invariant(s) are of 6'th
rather than 4'rth order, and $N_{inv}=N_{inv,6}=1$.  The phase constraint
yielded by this invariant is that the set of elements 
$\{A_{12}, A_{23}, A_{32}, A_{33}, B_{12}, B_{22} \}$ cannot be made 
simultaneously This allows $Y^{(u)}$ to be made real; if it is, then 
$\{B_{12},B_{22}\}$ cannot be made real.  The constraint also allows 
$Y^{(d)}$ to be made real, and if it is, then 
$\{A_{12},A_{23},A_{32},A_{33}\}$ cannot all be made real.

\subsection{Model 5}

   This model is given by 
\beq
Y^{(u)} =  \left (\begin{array}{ccc}
                  0 & 0 & A_{13} \\
                  0 & A_{22} & A_{23} \\
                  A_{31} & A_{32} & A_{33} \end{array}   \right  )
\label{yu5}
\eeq
and $Y^{(d)}$ as in (\ref{yd4}), 
\beq
Y^{(d)} =     \left ( \begin{array}{ccc}
                  0 &  B_{12} &  0 \\
                  B_{21} & B_{22} & 0 \\
                  0 & 0 & B_{33} \end{array}  \right )
\label{yd5}
\eeq
Here $N_{eq}=10$, and the $T$ matrix is $10 \times 9$ with 
rank 8, so that $N_p=2$.  We find two independent complex quartic
invariants: $P^{(u)}_{22,33}$ as in (\ref{pu2233}), and 
\beq
Q_{13,22}=A_{13}B_{22}A_{23}^*B_{12}^*
\label{q1322}
\eeq
Since the two complex quartic invariants have independent phases, they fully
account for all of the phase constraints.  These constraints are that 
it is not possible to
render the respective elements comprising the sets 
\beq
S_{P^{(u)}_{22,33}}=\{A_{22},A_{23},A_{32},A_{33}\}
\label{setpu2233}
\eeq
or
\beq
S_{Q_{13,22}}=\{A_{13},A_{23},B_{12},B_{22}\}
\label{setq1322}
\eeq
simultaneously real by any fermion rephasings.  These constraints allow one to
render $Y^{(d)}$ real.  If one does this, then one of the two unremovable 
phases must reside in the submatrix 
\beq
        \left ( \begin{array}{cc}
                  A_{22} & A_{23} \\
                  A_{32} & A_{33} \end{array}  \right )
\label{submatmod5}
\eeq
formed by the elements of set (\ref{setpu2233}) and the other in the set
$\{A_{13},A_{23}\}$.  The elements $A_{31}$, $B_{21}$, and $B_{33}$ do not
occur in any complex invariants and hence may be rephased freely. 
A possible form for the Yukawa matrices after rephasing is 
\beq
Y^{(u)'} =  \left ( \begin{array}{ccc}
                  0 & 0 & |A_{13}| e^{i \: \arg(Q_{13,22})} \\
     0 & |A_{22}|e^{i \: \arg(P^{(u)}_{22,33})} & |A_{23}| \\
         |A_{31}| & |A_{32}| & |A_{33}| \end{array}   \right  )
\label{yu5r}
\eeq
\beq
Y^{(d)'} =     \left ( \begin{array}{ccc}
                  0 &  |B_{12}| &  0 \\
                  |B_{21}| & |B_{22}| & 0 \\
                           0 & 0 & |B_{33}| \end{array}  \right )
\label{yd5r}
\eeq
Since the quartic complex invariants already contain all of the unremovable
phases, it follows from our theorem on invariants and its Corollary 1 that any
6'th order complex invariant yields no new phase constraint.  We find that
there is one 6'th order complex invariant, 
\beq
Q^{(uud)}_{32,13,22}=A_{32}A_{13}B_{22}A_{33}^*A_{22}^*B_{12}^*
\label{quud133222}
\eeq
Its argument is expressed in terms of those of quartic complex invariants as 
\beq
\arg(Q^{(uud)}_{32,13,22})= -\arg(P^{(u)}_{22,33})+\arg(Q_{12,22})
\label{q6mod5red}
\eeq
The model thus has $N_{inv,4}=2$, $N_{inv,6}=1$, and $N_{inv}=3$.

\subsection{Model 6}

   We have noted above that, to the extent that one understands the
phenomenological implications of string theories, they do not appear to yield
symmetric Yukawa matrices.  This is, indeed, part of the reason that we have
listed the previous five forms for these matrices in a form which is not, in
general, symmetric.  Here, as a further illustration of our methods, we
determine the unremovable phases and associated complex invariants for a
generalization of one specific model of Yukawa matrices inspired by a
particular 4D superstring construction and sketched in \cite{faraggi} 
\footnote{Ref. \cite{faraggi} does not give any discussion of the number of 
unremovable phases or which elements of the Yukawa matrices can be made real.}
The model is defined by 
\beq
Y^{(u)} =  \left (\begin{array}{ccc}
                  0 & A_{12} & 0 \\
                  A_{21} & A_{22} & A_{23} \\
                  0 & A_{32} & A_{33} \end{array}   \right  )
\label{yu6}
\eeq
\beq
Y^{(d)} =     \left ( \begin{array}{ccc}
                  0 & B_{12} &  0 \\
                  B_{21} & B_{22} & B_{23} \\
                  0 & B_{32} & B_{33} \end{array}  \right )
\label{yd6}
\eeq
As can be seen, if one takes $A_{22}=0$, this reduces to model 4.  The present
model has $N_{eq}=12$.  We calculate the $12 \times 9$ $T$ matrix as in the
previous cases and find that it has $rank(T)=8$, so that $N_p=4$.  We find that
there are $N_{inv,4}=7$ quartic complex invariants, two of $P_4$ type and five
of $Q_4$ type.  These are $P^{(u)}_{22,33}$, $P^{(d)}_{22,33}$, $Q_{12,22}$,
$Q_{12,32}$, $Q_{22,32}$, $Q_{22,33}$, $Q_{22,33}$, and $Q_{23,33}$.  The
ordering of the arguments of the complex quartic invariants in the vector 
$\xi_4$ may be taken to be the same as in the list just given. 
The corresponding $Z_4$ matrix for these order 4 invariants is the $N_{inv,4} 
\times N_{eq}=$ 7-row by 12-column matrix 
\beq
(Z_4)_6 =  \left (\begin{array}{cccccccccccc}
      0 & 0 & 1 &-1 &-1 & 1 & 0 & 0 & 0 & 0 & 0 & 0 \\
      0 & 0 & 0 & 0 & 0 & 0 & 0 & 0 & 1 &-1 &-1 & 1 \\
      1 & 0 &-1 & 0 & 0 & 0 &-1 & 0 & 1 & 0 & 0 & 0 \\
      1 & 0 & 0 & 0 &-1 & 0 &-1 & 0 & 0 & 0 & 1 & 0 \\
      0 & 0 & 1 & 0 &-1 & 0 & 0 & 0 &-1 & 0 & 1 & 0 \\
      0 & 0 & 1 & 0 &-1 & 0 & 0 & 0 & 0 &-1 & 0 & 1 \\
      0 & 0 & 0 & 1 & 0 &-1 & 0 & 0 & 0 &-1 & 0 & 1 \end{array} \right )
\label{z46}
\eeq
(where the second subscript refers to model 6).  We find that
$rank((Z_4)_6)=4$, so that the quartic complex invariants fully account for all
of the $N_p=N_{ia}=4$ unremovable phases.  Removing the last three rows of
$(Z_4)_6$ does not reduce the rank, which shows that the arguments of the 
first four quartic complex invariants in $(\xi_4)_6$ are linearly
independent, so that we may take these complex invariants, 
$P^{(u)}_{22,33}$, $P^{(d)}_{22,33}$, $Q_{12,22}$, and $Q_{12,32}$, as the set
with independent arguments.  Each of these complex invariants implies a phase
constraint that the corresponding set of elements cannot be made simultaneously
real.  A possible form of the Yukawa matrices, after rephasing, which is
consistent with the phase constraints, is 
\beq
Y^{(u)'} =  \left (\begin{array}{ccc}
         0 & |A_{12}|e^{i \: \arg(Q_{12,22})} & 0 \\
 |A_{21}| & |A_{22}| & |A_{23}|e^{i(- \arg(P^{(u)}_{22,33})
- \arg(Q_{12,22})+\arg(Q_{12,32}))} \\
 0 & |A_{32}|e^{i(\arg(Q_{12,22})-\arg(Q_{12,32}))} & |A_{33}| 
\end{array}   \right  )
\label{yu6r}
\eeq
\beq
Y^{(d)'} =     \left ( \begin{array}{ccc}
                  0 & |B_{12}| &  0 \\
        |B_{21}| & |B_{22}| & |B_{23}|e^{-i \: \arg(P^{(d)}_{22,33})} \\
            0 & |B_{32}| & |B_{33}| \end{array}  \right )
\label{yd6r}
\eeq
The model has $N_{inv,6}=4$ independent 6'th order complex invariants: 
\beq
Q^{(uud)}_{22,33,12}=A_{22}A_{33}B_{12}A_{23}^*A_{12}^*B_{32}^*
\label{quud223312}
\eeq
\beq
Q^{(uud)}_{33,12,22}=A_{33}A_{12}B_{22}A_{32}^*A_{23}^*B_{12}^*
\label{quud331222}
\eeq
$Q^{(ddu)}_{22,33,12}$ as given explicitly before in (\ref{qddu223312}) for
model 2, and 
\beq
Q^{(ddu)}_{33,12,22}=B_{33}B_{12}A_{22}B_{32}^*B_{23}^*A_{12}^*
\label{qddu331222}
\eeq
Thus, $N_{inv}=N_{inv,4}+N_{inv,6}=11$ for this model.  By our theorem on
invariants and its Corollary 1, the arguments of all of these 6'th order
complex invariants are expressible in terms of those of the four quartic 
complex invariants with independent arguments.  The methods have been fully
illustrated in our previous models, so we omit the explicit expressions.

\subsection{Model 7}

   We proceed to mention two toy models briefly to illustrate certain 
theoretical points. We have exhibited a special case of 
model 4 for which the only complex invariant occurs at 6'th order and is of
$Q_6$ type.  It is of interest to exhibit a form for the Yukawa matrices 
which again has only a 6'th
order complex invariant, but for which this is of $P_6$ rather than $Q_6$ 
type.  We have constructed such a model:
\beq
Y^{(u)} =  \left (\begin{array}{ccc}
                  0 & A_{12} & 0 \\
                  A_{21} & 0 & 0 \\
                  0 & 0  &   A_{33} \end{array}   \right  )
\label{yu7}
\eeq
and $Y^{(d)}$ as in (\ref{yd4}), 
\beq
Y^{(d)} =     \left ( \begin{array}{ccc}
                  B_{11} & B_{12} &  0 \\
                  B_{21} & 0 & B_{23} \\
                  0 & B_{32} & B_{33} \end{array}  \right )
\label{yd7}
\eeq
This model thus has $N_{eq}=9$.  We calculate the $T$ matrix and find that 
it has $rank(T)=8$, so that $N_p=1$.  By design, this model has
no quartic complex invariants, and a 6'th order complex invariant 
\beq
P^{(d)}_{11,23,32}=B_{11}B_{23}B_{32}B_{12}^*B_{21}^*B_{33}^*
\label{pd112332}
\eeq
The graphical representation of this invariant is given in Fig. 3(b) for $f=d$.
The elements of $Y^{(u)}$ are unconstrained by any phase invariants and, in
particular, $Y^{(u)}$ can be made real.  The phase constraint from 
$P^{(d)}_{11,23,32}$ forbids $Y^{(d)}$ from being made real or hermitian.  A
possible form for $Y^{(d)}$ after rephasing is
\beq
Y^{(d)'} =     \left ( \begin{array}{ccc}
                  |B_{11}| & |B_{12}| &  0 \\
                  |B_{21}| & 0 & |B_{23}|e^{i\: \arg(P^{(d)}_{11,23,32})} \\
                  0 & |B_{32}| & |B_{33}| \end{array}  \right )
\label{yd7r}
\eeq

\subsection{Model 8}

   We noted in section 2 that the rank of the $T$ matrix is 8 for all of the
realistic models that we have studied, but that, in principle, $rank(T)$ may be
less than 8.  We present here a toy model to illustrate this.  The model also
provides an illustration of the inequality (\ref{npnpv}), that the number of
unremovable phases in the Yukawa matrices is $\ge$ the number of phases in the
CKM quark mixing matrix, $V$.  Although formally the model has $N_G=3$, it is
of a degenerate type in which the third generation is decoupled from the 
first two in both the up and down quark sectors.  Hence, in some ways it acts
more like an $N_G=2$ model.  The Yukawa matrices are
\beq
Y^{(u)} =  \left (\begin{array}{ccc}
                  A_{11} & A_{12} & 0 \\
                  A_{21} & A_{22} & 0 \\
                  0      & 0      & A_{33} \end{array}   \right  )
\label{yu8}
\eeq
\smallskip
\beq
Y^{(d)} =     \left ( \begin{array}{ccc}
                  B_{11} & B_{12} &  0 \\
                  B_{21} & B_{22} & 0 \\
                  0      & 0      & B_{33} \end{array}  \right )
\label{yd8}
\eeq
So that $N_{eq}=10$.  The $T$ matrix is 
\beq
T_{8} =  \left (\begin{array}{ccccccccc}
     1 & 0 & 0 & 1 & 0 & 0 & 0 & 0 & 0 \\
     1 & 0 & 0 & 0 & 1 & 0 & 0 & 0 & 0 \\
     0 & 1 & 0 & 1 & 0 & 0 & 0 & 0 & 0 \\
     0 & 1 & 0 & 0 & 1 & 0 & 0 & 0 & 0 \\
     0 & 0 & 1 & 0 & 0 & 1 & 0 & 0 & 0 \\
     1 & 0 & 0 & 0 & 0 & 0 & 1 & 0 & 0 \\
     1 & 0 & 0 & 0 & 0 & 0 & 0 & 1 & 0 \\
     0 & 1 & 0 & 0 & 0 & 0 & 1 & 0 & 0 \\
     0 & 1 & 0 & 0 & 0 & 0 & 0 & 1 & 0 \\
     0 & 0 & 1 & 0 & 0 & 0 & 0 & 0 & 1 \end{array} \right )
\label{t8}
\eeq
with $rank(T_{8})=7$, so that $N_p=N_{eq}-rank(T)=3$.  The model has 
$N_{inv}=N_{inv,4}=6$ complex invariants, which are precisely those listed in 
eqs. (\ref{pf1122}) for $f=u$ and $d$ and in eqs. (\ref{q1122})-(\ref{q1222}) 
in section 4 for the $N_G=2$ model with all nonzero arbitrary elements.  
The $N_{ia}=N_p=3$ complex invariants among these with independent
arguments may be taken to be $P^{(u)}_{11,22}$, $P^{(d)}_{11,22}$, and
$Q_{11,22}$.  Note that this model also is like an $N_G=2$ model in that there
is no CP violation.  If one chooses to consider only the mutually coupled part
of the model, then the Yukawa matrices would reduce to the upper left $2 \times
2$ submatrices of (\ref{yu8}) and (\ref{yd8}), the model would 
reduce to a quasi $N_G=2$ model, and $T$ would become an $8 \times 9$ matrix 
with rank 5.  Hence for the mutually coupled sector of the model, the rank of 
$T$ would obey the relation that 
$rank(T)=rank(T)_{max}=3N_G-1$ (whence $rank(T)=5$ for $N_G=2$) 
analogous to the rank 8 nature of the true $N_G=3$ models.  When regarded as a
degenerate $N_G=3$ case, this model also constitutes an exception to the
empirical rule that $N_p = (N_p)_{max}-N_z$, since $(N_p)_{max}=10$ for $N_G=3$
and $N_z=8$.  However, again, if one considered only the mutually coupled
sector of the model, then it would be a quasi $N_G=2$ model, so that 
$(N_p)_{max}=3$, $N_z=0$, and the empirical relation would be obeyed.  Such
subtle distinctions are not necessary for the realistic models studied here.

\section{Conclusions}
\label{conclusions}

    The goal of understanding quark masses and quark mixing remains one of
the most important outstanding problems in particle physics.  When one
constructs a model of the quark Yukawa matrices, it is necessary to determine
how many unremovable phases there are in $Y^{(u))}$ and $Y^{(d)}$, and to
determine where these phases may be assigned, and which elements may be
rendered real by rephasings of fermion fields.  We previously reported a
general solution to this problem \cite{ks} involving several theorems and a
related analysis of complex rephasing-invariant products of elements of the
Yukawa matrices.  In the present paper we have given a detailed discussion of 
our methods and results.  We have also presented new results on phases and
complex invariants for the case of arbitrary $N_G$, the number of fermion
generations, since this elucidates their theoretical properties and gives
further insight into the case of physical interest, $N_G=3$.  Finally,
we have given a detailed application of our methods to currently viable models.

   This research was supported in part by NSF grant PHY-93-09888.  We thank L.
Lavoura for a discussion on the quartic $P$ invariant. 
\vfill
\eject

\vfill
\eject

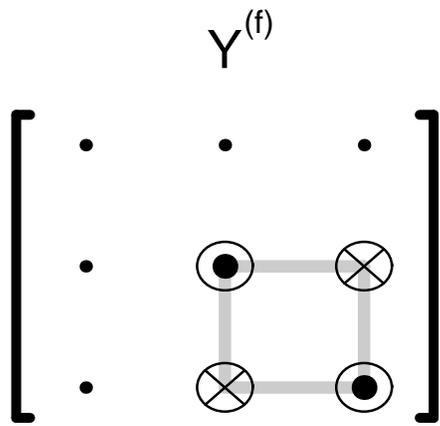
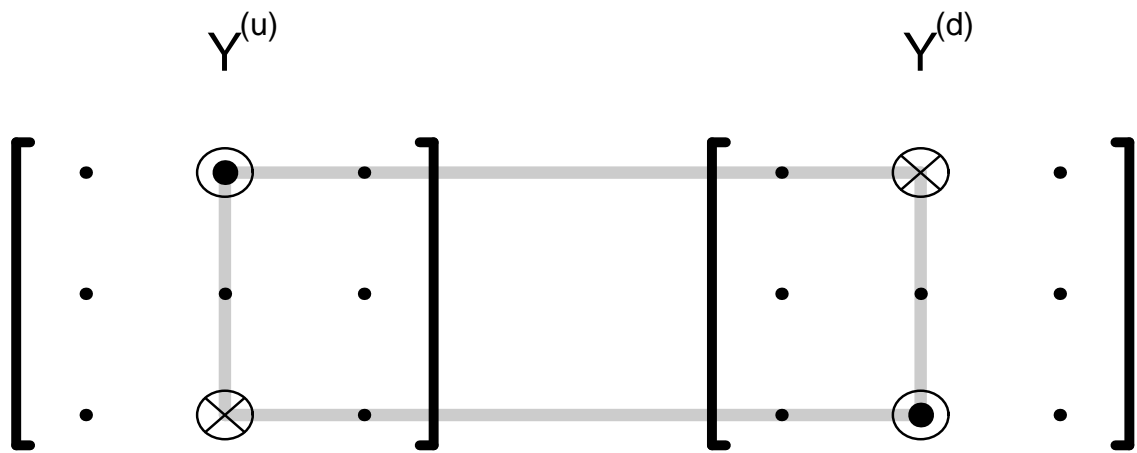
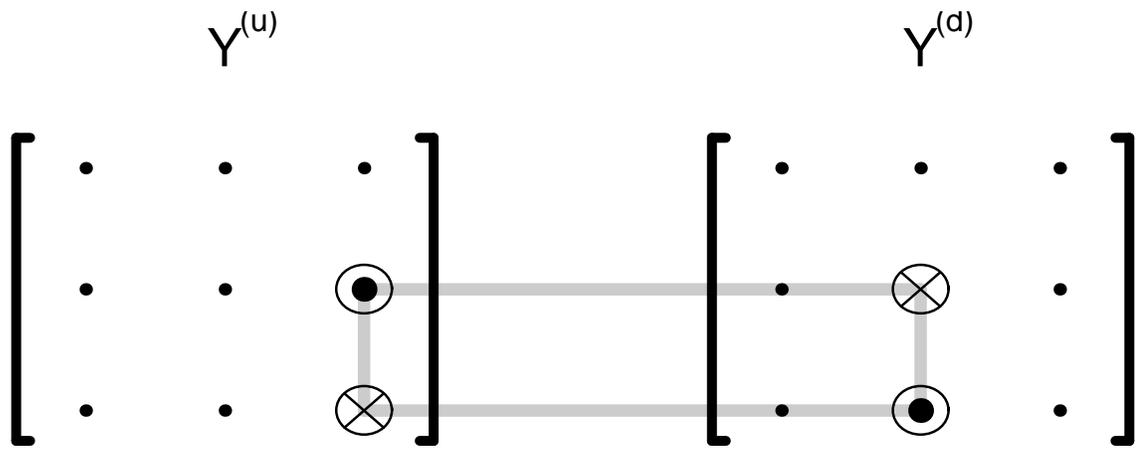
\end{document}